\documentclass[]{aastex631} 
\usepackage[caption=false]{subfig}
\usepackage{multirow}
\usepackage{graphicx}
\usepackage{booktabs}
\usepackage{amsmath,amssymb}
\usepackage[T1]{fontenc}
\usepackage{threeparttable}
\usepackage{url}
\begin{document}

\title{Event Rate Density and Luminosity Function of Newborn-Magnetar-Driven X-Ray Transients from Neutron Star Binary Mergers}
\correspondingauthor{En-Wei Liang}
\email{lew@gxu.edu.cn}
\author{Le Zou}
\affil{Department of Physics, Xiangtan University, Xiangtan 411105, China}
\affil{Key Laboratory of Stars and Interstellar Medium, Xiangtan University, Xiangtan 411105, China}
\author{Ji-Gui Cheng}
\affil{School of Physics and Electronics, Hunan University of Science and Technology, Xiangtan 411201, China}
\author{Rui-Chong Hu}
\affil{Guangxi Key Laboratory for Relativistic Astrophysics, School of Physical Science and Technology, Guangxi University, Nanning 530004, China}
\author{Wen-Jin Xie}
\affil{Université Paris Cité, CEA Paris-Saclay, IRFU/DAp-AIM, 91191 Gif-sur-Yvette, France}
\author{En-Wei Liang}
\affil{Guangxi Key Laboratory for Relativistic Astrophysics, School of Physical Science and Technology, Guangxi University, Nanning 530004, China}

\begin{abstract}
X-ray transients (XTs) driven by newborn magnetars from mergers of neutron star binaries (NSBs) were occasionally detected in the narrow-field {\it Chandra} Deep Field-South survey (CDF-S) and the {\it Swift}/XRT observations of short gamma-ray bursts (sGRBs). Quantifying their event rate density (ERD) and luminosity function (LF) is critical for understanding NSB coalescence and magnetar formation. Utilizing population synthesis calculations incorporating various equations of state (EoS), we derive a local ERD of $\sim 300\,{\rm Gpc^{-3}\,yr^{-1}}$ and a redshift-dependent ERD profile peaking at $z=1.81$ followed by rapid decline beyond $z \sim 4$. Constructing an XT sample based on CDF-S and {\it Swift} observations, we characterize the LF by a single power-law function at $L \leq 4.75 \times 10^{46}\;{\rm erg\;s^{-1}}$ with a slope of $-1.03$, following by a broken power-law function in which the break luminosity is $L_{\rm b} = 4.38 \times 10^{47}\;{\rm erg\;s^{-1}}$ and the slopes are $-0.28$ and $-1.66$. Based on the ERD and the LF, we estimate that the {\it Einstein Probe} ({\it EP}) detection rate is $\sim 31\;{\rm yr^{-1}}$, adopting a conservative threshold flux of $10^{-9}\;{\rm erg}\;{\rm s^{-1}}$, an luminosity range of $L \in [2\times 10^{44},2\times 10^{49}]\;{\rm erg\;s^{-1}}$, and a correction for jet opening angle of $\sim 16^{\circ}$. This detection rate is consistent with the {\it EP} observations during its first-year operation. It is important to note that our estimation is subject to uncertainties arising from the LF derivation. Future {\it EP} observations of these XT events will be crucial in reducing these uncertainties.
\end{abstract}


\section{Introduction}
The discovery of the association among gravitational wave (GW) burst, gamma-ray burst (GRB), and kilonova on Aug. 17, 2017, i.e. GW 170817/sGRB 170817A/AT2017gfo, from a neutron star binary (NSB) merger opens the multi-messenger astronomical era \citep{2017PhRvL.119p1101A, 2017ApJ...848L..12A}. It is highly prospective that NSB mergers are the sites for the synergic observations of the multi-wavelength electromagnetic (EM) and GW signals. Theoretical studies and simulations suggest that an NSB merger can give birth to either a black hole (BH) or a neutron star (NS), depending on the total mass of the binary system and the NS equation of state (EoS; \citealt{2010MNRAS.409..531R, 2013ApJ...771L..26G, 2014PhRvD..89d7302L, 2016PhRvD..94h3010L}). The newborn NSs could be classified into three different types, i.e. a hyper-massive NS (HMNS), which survives for only about $100$ ms before collapsing into a BH \citep{2000ApJ...528L..29B, 2006PhRvD..73f4027S}, a supra-massive NS (SMNS), which has a longer lifetime ranging from a few seconds to several hours before collapse \citep{2010MNRAS.409..531R, 2015ApJ...805...89L, 2016PhRvD..93d4065G}, and a stable NS \citep{1998A&A...333L..87D, 2001ApJ...552L..35Z, 2010ApJ...715..477Y, 2011MNRAS.413.2031M}. It is suggested that all three types of NS possess a strong magnetic field (i.e. $10^{14} \sim 10^{15}\,{\rm G}$; \citealp{2013ApJ...771L..26G, 2015ApJ...809...39G}) and can be recognized as magnetars \citep{1992ApJ...392L...9D, 1995MNRAS.275..255T, 1996ApJ...473..322T}.

It has been suggested that newborn magnetars are powerful central engines of GRBs \citep{2007ApJ...665..599T,2013MNRAS.430.1061R, 2015ApJ...805...89L}. These magnetars dissipate their spin energy via the GW emission and magnetic dipole radiation (DR). The continuous energy injection from the DR can power X-ray transients (XTs), in addition to the GRBs and their afterglows as well as the kilonova. Their lightcurves are typically characterized as an X-ray plateau or a shallow decay segment, i.e. $L\propto(1+t/\tau)^{-2}$, where $\tau$ is the characteristic spin-down timescale of a magnetar \citep{1992Natur.357..472U, 1998A&A...333L..87D, 2001ApJ...552L..35Z, 2011MNRAS.413.2031M, 2014ApJ...785...74L}. Such a signature is not only present in early X-ray afterglow lightcurves of long GRBs (lGRBs; \citealt{2007ApJ...665..599T}) but also is observed in the X-ray lightcurves of some short GRBs (sGRB; \citealt{2013MNRAS.430.1061R, 2015ApJ...805...89L}). Thus, magnetar-driven XTs make diversity of the observed early X-ray afterglow lightcurves of GRBs, depending on the relative flux levels between the fireball afterglows and the XTs \citep{2018MNRAS.480.4402L}. The XTs may be identified through X-ray follow-up observations (e.g., in GRBs 070110A and 180620A; \citealt{2007ApJ...665..599T,2022MNRAS.513L..89Z}) and are occasionally detected as orphan X-ray or soft gamma-ray transients. As proposed by \cite{2021ApJ...921L...1Z}, the peculiar GRB 101225A detected with {\it Swift}/BAT is a bright orphan XT powered by the DR of a newborn magnetar. More interestingly, \cite{2019Natur.568..198X} found that the X-ray lightcurve of the redshift-known CDF-S XT2 is consistent with predictions of XTs driven by newborn magnetars. Several orphan redshift-known XTs have already been discovered in the {\it Chandra} deep-field X-ray survey (CDF-S) data \citep{2022A&A...663A.168Q, 2023A&A...675A..44Q}.

Magnetar-driven XTs are crucial probes for investigating NS physics. Based on the formation history of magnetars, these XTs can be further divided into two groups, i.e. ones powered by magnetars formed in NSB mergers (NSB-magnetars) and ones powered by magnetars born from the collapse of massive stars (MS-magnetars). Over the years, many researches have been conducted to determine the luminosity function (LF) and the event rate density (ERD) of the two XTs populations. \cite{2020ApJ...894...52X} constrained the LF of the MS-magnetar-driven XTs as a smooth broken power-law breaking at $L_{\rm b}=3.2\times 10^{48}$ erg s$^{-1}$ with index changing from $\beta_{1}=-0.79$ to $\beta_{2}=-2.22$. \cite{2017ApJ...835....7S} constrained the ERD and the LF of the NSB-magnetar-driven XTs by adopting a redshift distribution of NSB mergers derived from empirical merger delay model \citep{2011ApJ...727..109V,2015MNRAS.448.3026W}. Their results showed that the LF can be fitted with a two-log-normal distribution peaking around $2.5\times 10^{46}\;{\rm erg\;s^{-1}}$ and $4.0\times 10^{49}\;{\rm erg\;s^{-1}}$, and the ERD is around a few tens of Gpc$^{-3}\;{\rm yr^{-1}}$ for luminosity above $10^{45}\;{\rm erg\;s^{-1}}$. Despite these studies, the ERD and LF of NSB-magnetar-driven XTs remain poorly understood, due to the absence of wide field-of-view (FoV) X-ray surveys. 

The launch of the soft X-ray {\it Einstein Probe} ({\it EP}) telescope on January 9, 2024, presents an excellent opportunity to investigate NSB-magnetar-driven XTs. In its first year of operation, {\it EP} has detected 69 XTs, including EP240315a \citep{2024GCN.35931....1Z}, EP240408a \citep{2024GCN.36053....1H}, EP240413a \citep{2024GCN.36086....1L}, and EP240414a \citep{2024GCN.36091....1L}. Most of these detections exhibit peak fluxes around $1 \times 10^{-9}\;{\rm erg\;cm^{-2}\;s^{-1}}$, with a few reaching up to $1 \times 10^{-8}\;{\rm erg\;cm^{-2}\;s^{-1}}$. Some of detections may relate to NSB-magnetars. GW observations are also providing valuable insights, as NSB-driven XTs are among the most promising EM counterparts of GW sources. For instance, \cite{2019PhRvX...9c1040A} infer a local NSB merger rate density of $110-3840\;{\rm Gpc^{-3}\;yr^{-1}}$ at a 90\% confidence interval, using the O1/O2 runs of Advanced LIGO and Virgo. This merger rate can be used to constrain the ERD of NSB-driven XTs. In this paper, we aim to estimate the {\it EP} detection rate of NSB-magnetar-driven XTs and compare it with the {\it EP} observations. In Sec. \ref{sec:R(z)}, we use population synthesis simulations, combined with current GW observations, to obtain the ERD. In Sec. \ref{sec:LF}, we derive the LF via empirical function fitting of the {\it Swift} and {\it Chandra} sample of NSB-magnetar-driven XTs candidates. The final {\it EP} detection rate is presented in Sec. \ref{sec:EP}. Discussions and conclusions are given in Sec. \ref{sec:Dis} and Sec. \ref{sec:Con}, respectively. Throughout this paper, we adopt the standard $\Lambda$CDM cosmology with parameters $H_0=70.0,{\rm km,s^{-1},Mpc^{-1}}$, $\Omega_{\rm M} = 0.3$, and $\Omega_\Lambda = 0.7$.

\section{Event Rate density}
\label{sec:R(z)}
We investigate the ERD of NSB-magnetars based on available results of population synthesis simulations implemented by the {\tt StarTrack} code, which is used to generate populations of NSB systems \citep{2002ApJ...572..407B, 2008ApJS..174..223B}. This code has been improved by \cite{2020A&A...636A.104B}, including updated treatments of a common envelope (CE) evolution, compact object mass calculations, the new stellar remnant natal kick, and the angular momentum transport. \cite{2020A&A...636A.104B} generated a broad set of submodels of compact binary systems using the upgraded code. We adopt the results of the submodel M33.A for our analysis since predictions of this model are consistent with the observational limits for generating the population of NSBs. In this model, the rapid supernova engine model for NS mass is adopted from \cite{2012ApJ...749...91F} with a high natal kick ($\sigma= 265$  km $\rm s^{-1}$). 50\% non-conservative Roche lobe overflow and 5\% Bondi-Hoyle accretion are set for the CE evolution modeling. The cosmic star formation density is taken from \cite{2017ApJ...840...39M}. The average metallicity is $Z=0.014 Z_{\odot}$, where $Z_{\odot}$ is the solar metallicity. And for more details of the simulations please refer to \cite{2020A&A...636A.104B}.

Based on the simulation results, we are allowed to track every NSB event. Their merger information, including the total mass of the binary system $M_{\rm tot}$ and the merger delay timescale, is recorded. The typical mass of material outflow from an NSB system is $\sim 10^{-4}-10^{-2} M_{\odot}$ (e.g. \citealt{2011ApJ...732L...6R, 2013MNRAS.430.2585R, 2013PhRvD..87b4001H}), which could be negligible in comparison with the total mass of the NSB system ($M_{\rm tot}$). Therefore, the merger remnant mass can be approximated as $M_{\rm tot}$. $M_{\rm tot}$ of an NSB system places a strong constraint on the outcome of the remnant, which could be classified into four types concerning the NS maximum mass $M_{\rm TOV}$ (the Tolman-Oppenheimer-Volkov mass), i.e. a BH with $M_{\rm tot}/M_{\rm TOV} \gtrsim 1.3-1.6$ \citep{1994ApJ...424..823C}, an HMNS with $\sim 1.2 \lesssim M_{\rm tot}/M_{\rm TOV} \lesssim 1.3-1.6$, an SMNS with $1\lesssim M_{\rm tot}/M_{\rm TOV} \lesssim 1.2$, and a stable NS with $M_{\rm tot}/M_{\rm TOV} < 1$ \citep{2017ApJ...850L..19M}. Since we consider only the remnant as an NS, we exclude NSB events with $M_{\rm tot} \gtrsim 1.3 M_{\rm TOV}$. The $M_{\rm TOV}$ depends on the EoS of an NS.

\begin{figure}
\center
\includegraphics[angle=0,scale=0.35]{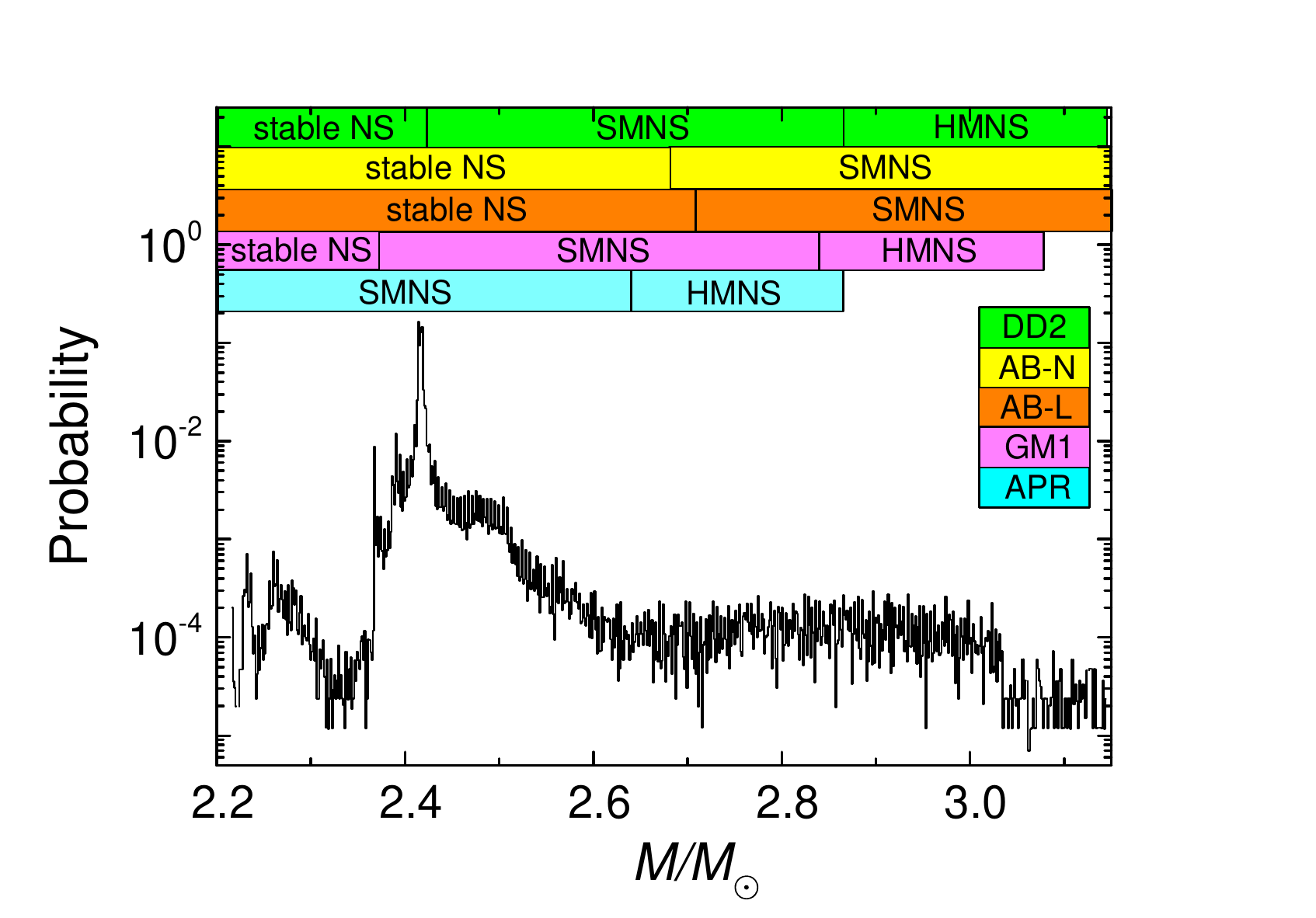}
\caption{The distribution of the remnant masses of NSB mergers derived from the population synthesis simulations implemented by the {\tt StarTrack} code. According to the different EoSs with the $M_{\rm TOV}$, we classify the remnant masses into the different types, i.e. an HMNS with $\sim 1.2 M_{\rm TOV}\lesssim M_{\rm tot} \lesssim 1.3-1.6 M_{\rm TOV}$, an SMNS with $M_{\rm TOV}\lesssim M_{\rm tot} \lesssim 1.2 M_{\rm TOV}$, and a stable NS with $M_{\rm tot} < M_{\rm TOV}$ \citep{2017ApJ...850L..19M}. The different EoSs are marked with color bars. The magenta, green, yellow, orange, and cyan present the EoSs GM1, DD2, AB-N, AB-L, and APR, respectively.}
\label{fig:m}
\end{figure}

The distribution of the $M_{\rm tot}$ for the selected remnants of NSB is illustrated in Fig. \ref{fig:m}. The range of $M_{\rm tot}$ spans from $2.22M_{\odot}$ to $3.15M_{\odot}$. We exclude EoSs that result in a $M_{\rm TOV}$ lower than $2.22M_{\odot}$. For our analysis, we adopt four EoSs with the following corresponding $M_{\rm TOV}$ values: EoS DD2 with $M_{\rm TOV}=2.42M_{\odot}$ \citep{2010PhRvC..81a5803T}, EoS AB-L with $M_{\rm TOV}=2.71M_{\odot}$, EoS AB-N with $M_{\rm TOV}=2.67M_{\odot}$ \citep{1977ApJS...33..415A}, and EoS GM1 with $M_{\rm TOV}=2.37M_{\odot}$ \citep{1991PhRvL..67.2414G}. It is should be noted that EoS APR with $M_{\rm TOV}=2.20M_{\odot}$ \citep{1998PhRvC..58.1804A}, which is only slightly smaller than $2.22M_{\odot}$. Hence, the EoS APR is also considered in our analysis. The mass ranges of the different remnant types, determined by the $M_{\rm TOV}$ corresponding to each EoS, are indicated by color bars in Fig. \ref{fig:m}.

\begin{figure}
\center
\includegraphics[angle=0,scale=0.3]{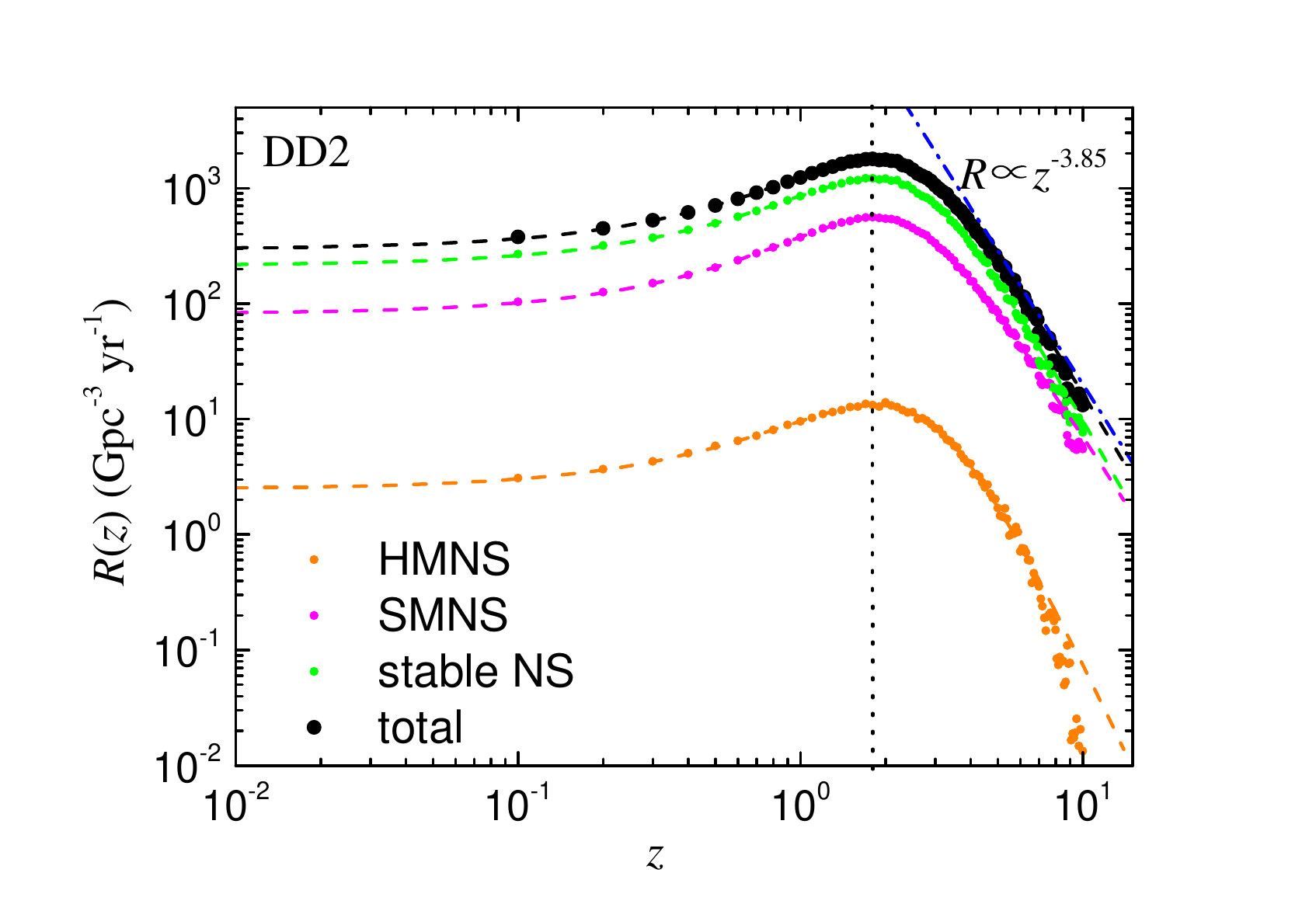}\\
\includegraphics[angle=0,scale=0.3]{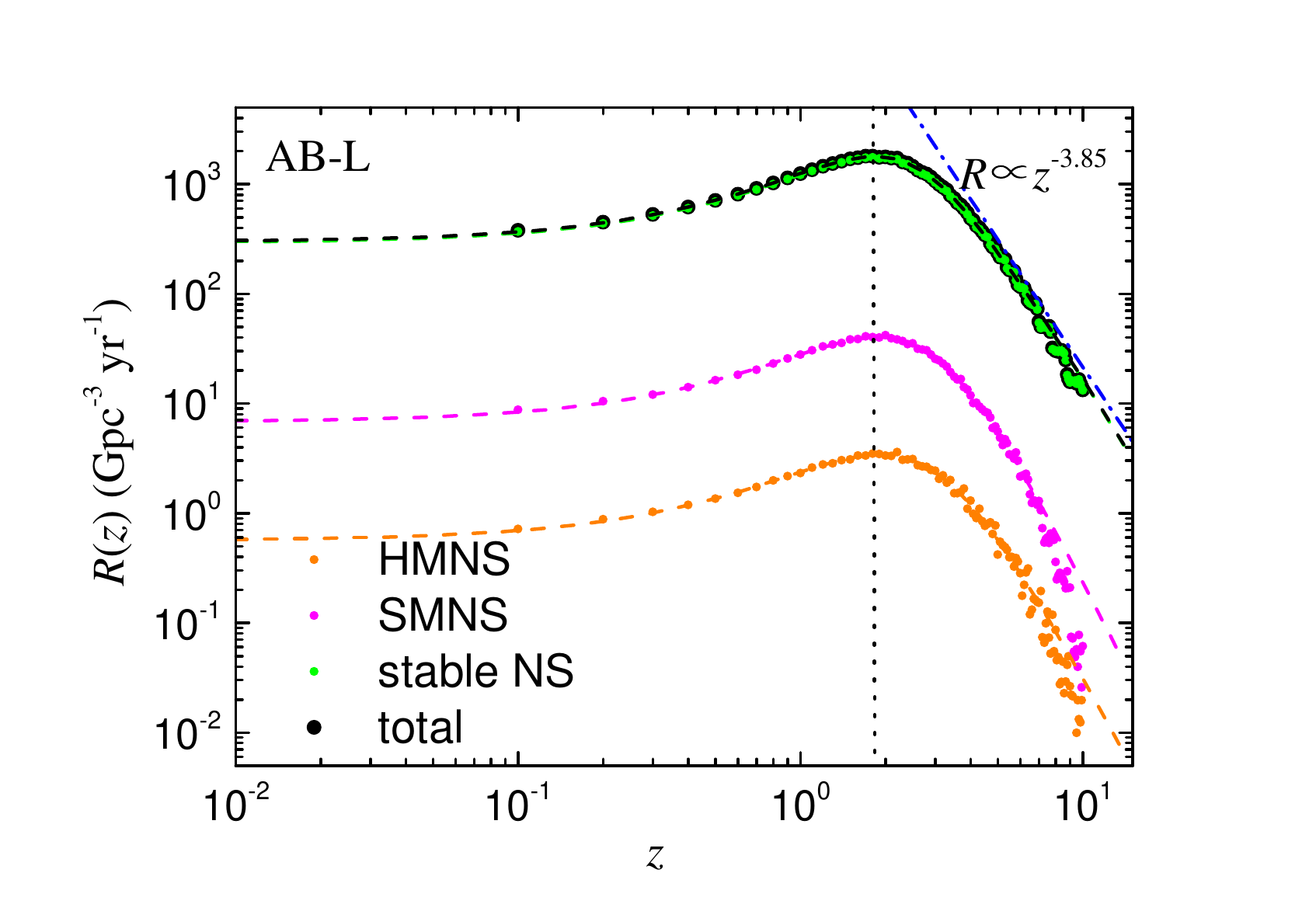}
\includegraphics[angle=0,scale=0.3]{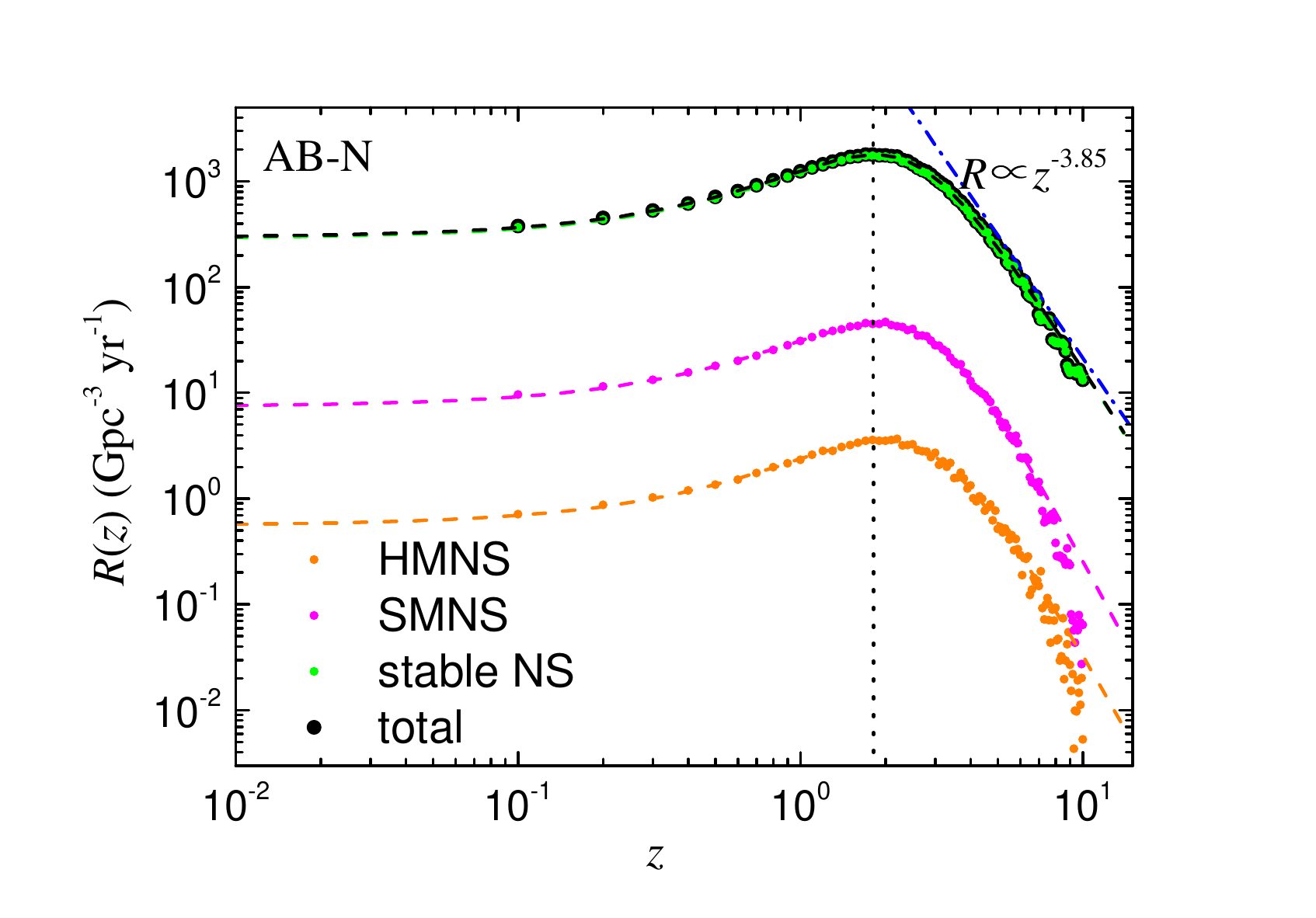}
\includegraphics[angle=0,scale=0.3]{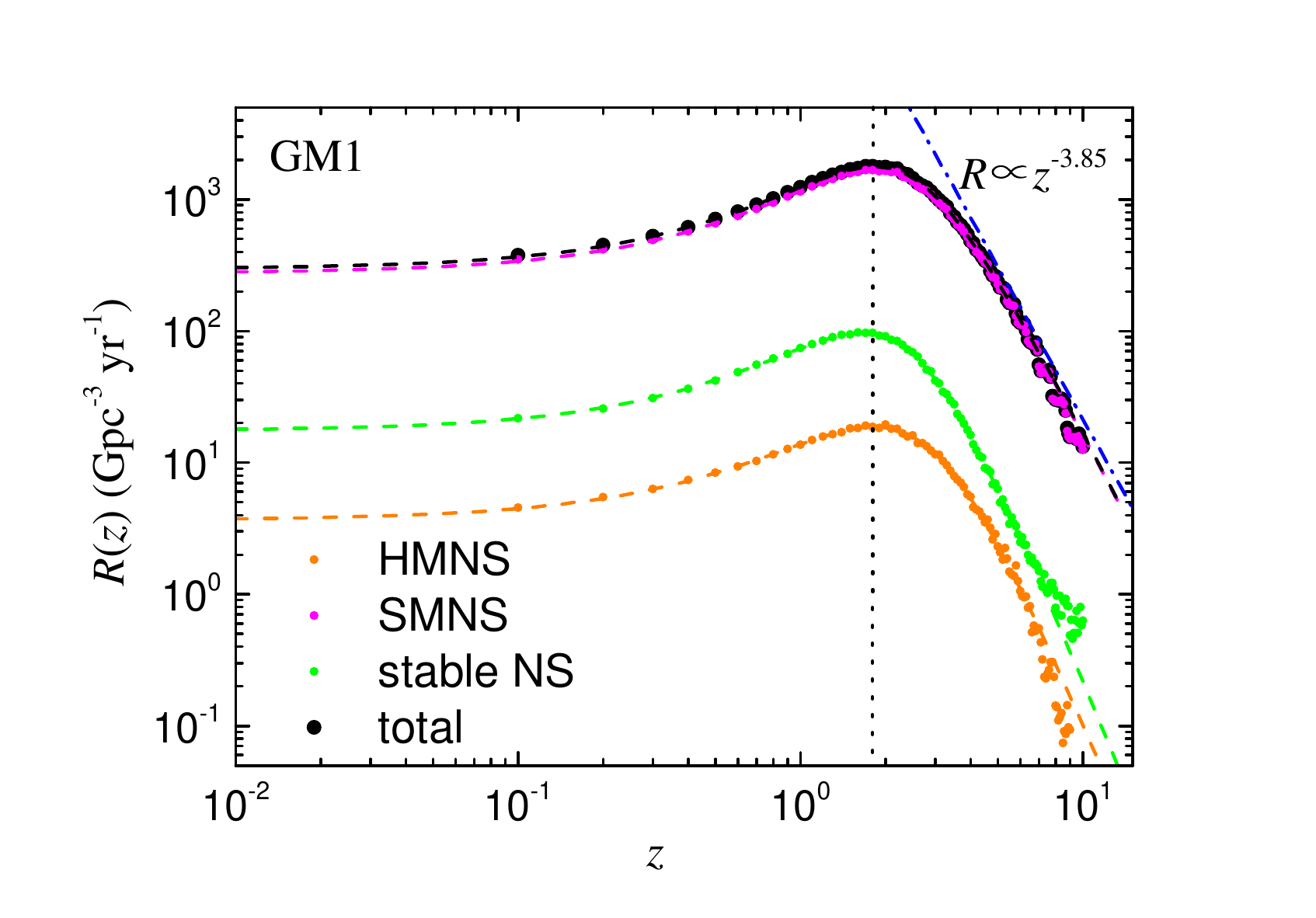}
\includegraphics[angle=0,scale=0.3]{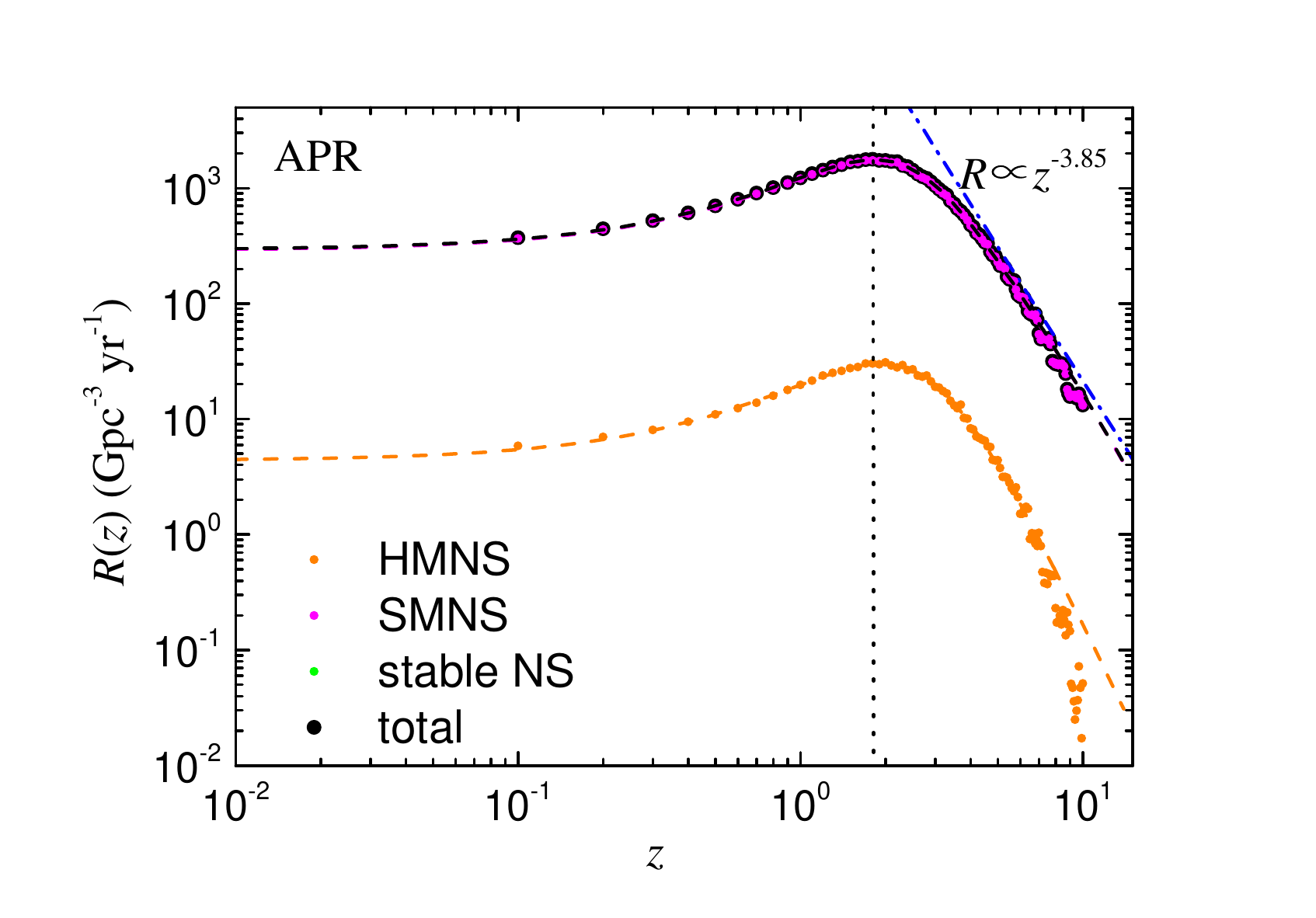}
\caption{Cosmic event rate history of the NSB-magnetar-XTs derived from \texttt{StarTrack} M33.A population synthesis simulation with different EoSs. The black dots are the total simulation data, green, magenta, and orange dots present the redshift distributions of different remnants, respectively. The black, green, magenta and orange dashed lines are the best-fit empirical model given in Eq.\ref{eq:R(z)}. The black dot line presents the peak $R$ at $z=1.81$. The blue dashed dot line marks the cosmic evolution $R\propto z^{-3.85}$ at $z> 4$.}
\label{fig:R(z)}
\end{figure}

The NSB ERDs ($R(z)$) as a function of $z$ for the global sample, the stable NS sample, the SMNS sample, and the HMNS sample are shown in Fig. \ref{fig:R(z)}. We fit these curves with an empirical model, i.e.
\begin{equation}
	\label{eq:R(z)}
	R(z) = R_{\rm 0} \left[ \left(1+z\right)^{a \eta} + \left( \frac{1+z}{B} \right)^{b \eta} \right]^{\frac{1}{\eta}},
\end{equation}
where $R_{0}$ approximates the local ERD at $z=0$ and $\eta$ measures the curvedness of the curve. This model is composed of two smoothly-joint broken power-law functions. The parameter $B$ is related to the slopes of the two components as $B=(1+z_{\rm b})^{1-a/b}$, where $z_b$ is the transition redshift of the two components. It has an advantage for the derivation of the local ERD. We employ a least-square algorithm to fit the curves to the data and the results of fitting are summarized in Tab. \ref{tab:fit} and depicted in Fig. \ref{fig:R(z)}. Our results demonstrate that the model can well fit the data at $z<5$. However, at $z>5$, the ERDs of SMNS and HMNS samples deviate from the fitting curves for certain EoSs. The derived parameter values, namely $a$, $b$, and $B$, are roughly consistent among the NS samples in different EoSs. Globally, the local total ERD is $R_{\rm 0}\sim300\,{\rm Gpc^{-3}\,\rm\,yr^{-1}}$, with a peak ERD $R_{\rm peak}$ occurring at $z_{\rm peak} = 1.81$. Beyond $z=4$, $R(z)$ drops rapidly as $R\propto z^{-3.85}$. \cite{2022MNRAS.509.1557C} investigated the properties of NSBs and their mergers by combining population synthesis models for binary stellar evolution (BSE) with cosmological galaxy formation and evolution models. Their methodology is consistent with that of in \cite{2020A&A...636A.104B}, albeit with different initial and boundary conditions. By incorporating observed galactic NSBs and the local NSB merger rate density inferred from GW observations, they constrained the NSB merger rate density as $R(z)\sim R_{0}(1+z)^{\zeta}$ at $z\lesssim 0.5$, where $R_{0} \sim 316-784\,{\rm Gpc^{-3}\,\rm\,yr^{-1}}$ and $\zeta \sim 1.34-2.04$.

The values of $R_0$ differ significantly among the remnant types for various EoSs. In the scenarios of EoSs DD2, AB-L, and AB-N scenarios, the event rate of the stable NSs dominates the total event rate, especially in the latter two scenarios. Conversely, in the EoS GM1 and APR scenarios, the event rate of SMNSs dominates the global sample due to the relatively low $M_{\rm TOV}$ values. Even no stable NS can be produced in the EoS APR scenario.

\section{Luminosity Function}
\label{sec:LF}

At present, no wide FoV survey sensitive in the soft X-ray band is available for constraining the luminosity function (LF) of the XTs. Some of the XTs are occasionally detected with the {\it Swift} and {\it Chandra} missions. Early X-ray follow-up observations with the {\it Swift}/XRT reveal a steady X-ray emission component embedded in the jet afterglow fireball of about half of the GRBs triggered by the {\it Swift}/BAT. We attribute this component to an XT powered by the DRs of a newborn magnetar. Considerable samples of the XTs are accumulated with the XRT data over $\sim 20\;{\rm yr}$ operation of the {\it Swift}/XRT emission. Moreover, the extended soft gamma-rays observed in some GRBs with the {\it Swift}/BAT may also be from the XTs \citep{2013ApJ...763L..22Z, 2015PhR...561....1K}.

We conduct a comprehensive analysis of XTs and sGRBs reported in the literature, which are believed to originate from NSB-magnetars. \cite{2021MNRAS.508.2505Z} analyzed the {\it Swift} GRB catalog up to 2021, identifying 28 sGRBs with X-ray plateaus (XPs), of which 13 had measured redshifts. This subset accounted for $\sim 26 \%$ of all sGRBs recorded at that time. By analyzing the {\it Swift} GRB catalog after 2021, we find only one additional sGRB with an XP but without a redshift measurement, indicating that sGRBs with XPs currently represent $\sim 21 \%$ of the {\it Swift} sGRBs catalog, with only $\sim 10 \%$ having both XPs and redshift measurements. We also select 11 extended emissions (EEs) with redshift measurements in sGRBs as reported by \cite{2014ApJ...783...24L}, \cite{2015ApJ...805...89L}, \cite{2018MNRAS.481.4332A}, and \cite{2021ApJS..252...16L}, and collect 10 plausible XTs with a plateau phase from \cite{2017MNRAS.467.4841B}, \cite{2019Natur.568..198X}, \cite{2022A&A...663A.168Q}, and \cite{2023A&A...675A..44Q}. Considering the luminosity of some XTs falls below $10^{44}$ erg $\rm s^{-1}$, which is unlikely to be produced by a NSB-magnetar \citep{2014MNRAS.439.3916M,2017ApJ...835....7S,2018ApJ...857...95M}, we exclude 5 XTs from our sample. Finally, these data allowed us to compile a sample of 29 events.

For these events, we model their lightcurves with an empirical broken power-law function, i.e. $F = F_{\rm 0} \left[(t / t_{\rm b})^{\omega\alpha_1} + (t / t_{\rm b})^{\omega\alpha_2}\right]^{-1/\omega}$, in which $t_{\rm b}$ presents the break time, $\omega$ describes the sharpness of the break, $\alpha_1$ and $\alpha_2$ are the decay indices before and after $t_{\rm b}$, respectively. To determine the photon spectral index ($\Gamma$), we fit the spectra of the magnetic dipole emission-dominated epoch using an absorbed single power-law function. The parameters of our sample are detailed in Tab. \ref{tab:paras}, where $F_{\rm b}$ corresponding to the break flux at $t_{\rm b}$ and $L_{\rm b}=4\pi D_{L}^2 F_{\rm b}$ representing the corresponding luminosity. We adopt the break luminosity as the characteristic luminosity for our sample.

\begin{figure}
\center
\includegraphics[angle=0,scale=0.30]{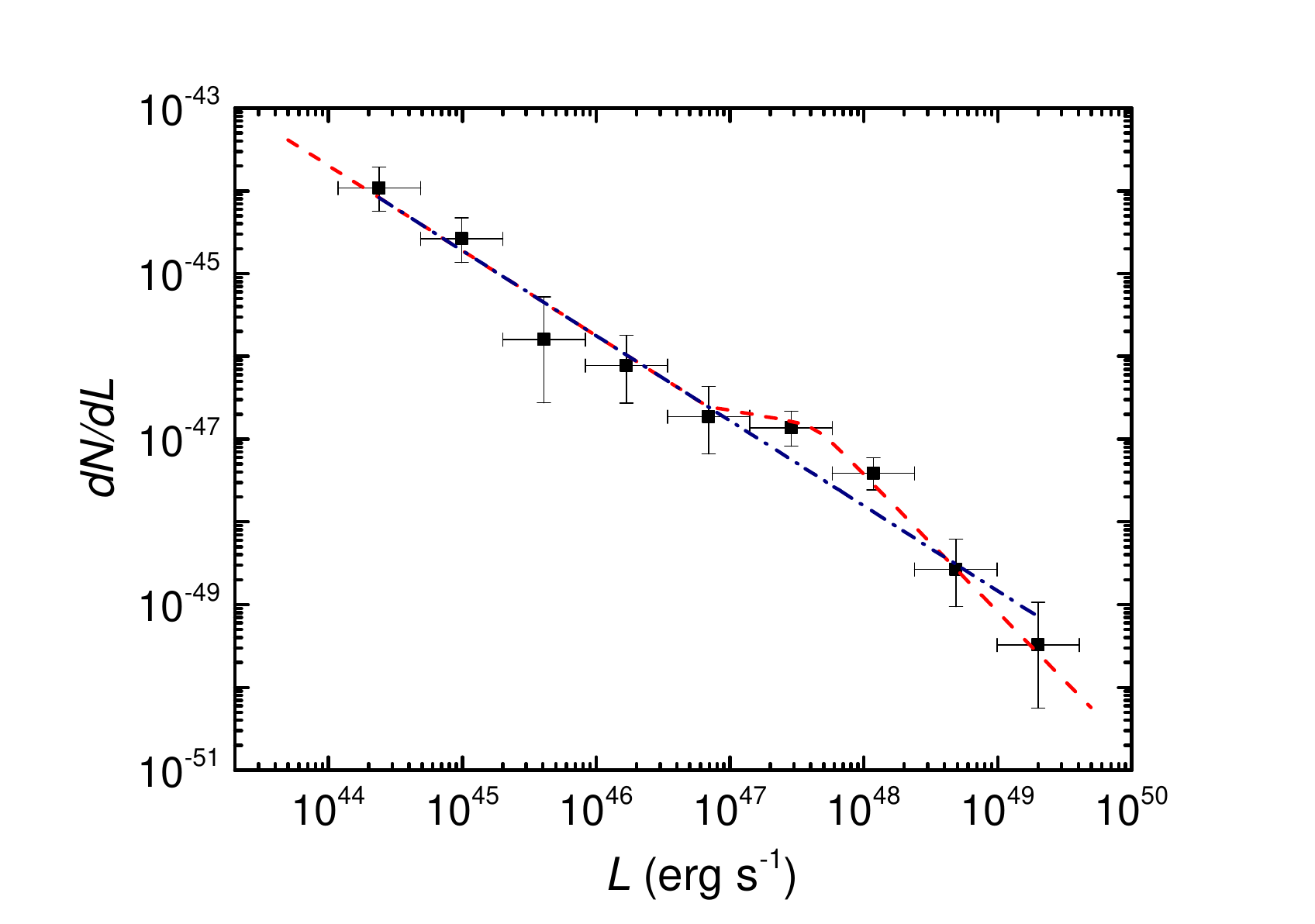}
\caption{The constructed luminosity function of the XTs driven by NSB-magnetars. The blue dashed dot line and red dashed line are the best-fit single power-law function and a combination of a single power-law function and a broken power-law function, respectively. The data with error bars are calculated with the XTs reported in Table 2.
}
\label{fig:LF}
\end{figure}

In Fig. \ref{fig:LF}, we categorize the XTs observed by different instruments into 9 equal-interval luminosity bins in the logarithmic space. We initially applied a single power-law function to fit the data. The result shows that the function does not fit well after the fifth point data. Consequently, we consider employing a combination of a single power-law function and a broken power-law function, i.e. $\Phi(L)\propto  L^{a}$ for data points before the fifth and $\Phi(L)\propto  [(L/L_{\rm b})^{\omega\beta_1}+ (L/L_{\rm b})^{\omega\beta_2}]^{1/\omega}$ for the subsequent points, to better constrain the LF. The best fit yields $a=-1.03$, $\beta_{1}=-0.28$, $\beta_{2}=-1.66$, and $L_{\rm b}=4.38\times 10^{47}$ $\rm erg \rm s^{-1}$. Our results show that the derived LF aligns well with the observed data, as illustrated in Fig. \ref{fig:LF}. Note that the derived LF is expected to have uncertainties arising from two factors. One is the instrument selection effect in constructing the XT sample. The {\it Chandra} just occasionally captures some XTs in the deep survey and the {\it Swift}/XRT only tracks the rapid follow-up of GRB afterglows, neither of which is a all-sky survey project. And the other is the relatively small sample size. These factors may cause the derived LF to deviate significantly from the true one. However, such uncertainties are currently unavoidable, as the derivation of the LF relies on the available observational sample. In the future, as the number of detected XTs increases, these uncertainties will be significantly mitigated.

\section{Prospect of the {\it EP} XT Survey}
\label{sec:EP}

For a detector with a flux threshold of $F_{\rm th}$ and an FoV of $\Omega$, the number of detectable events within an operational time $T$ can be estimated through
\begin{equation}
	\label{eq:dNdL}
	N = \frac{\Omega T}{4 \pi} \int_{L_{\rm min}}^{L_{\rm max}} \Phi(L) dL \int_{0}^{z_{\rm max}(L)} \frac{R(z)}{1+z} \frac{dV(z)}{dz} dz,
\end{equation}
where $\Phi(L)$ is normalized to the unit in luminosity range [$L_{\rm min}$, $L_{\rm max}$], $z_{\rm max}$ is the maximum detectable redshift for an XT with luminosity $L$, and $dV/dz$ is the co-moving volume at redshift $z$, which is calculated under the standard $\Lambda$CMD cosmology as
\begin{equation}
    \label{eq:dV}
	\frac{dV}{dz} = \frac{c}{H_{\rm 0}} \frac{4 \pi D_{\rm L}^{2}}{(1 + z)^{2} [\Omega_{\rm M}(1 + z)^{3} + \Omega_{\rm \Lambda}]}.
\end{equation}
The $z_{\rm max}$ is determined by the threshold flux through
\begin{equation}
    \label{eq:F}
	F_{\rm th} = \frac{L}{4 \pi D_{\rm L}^{2}(z_{\rm max}) k(z_{\rm max}, \Gamma)},
\end{equation}
where $k(z, \Gamma)$ is the $k$-correction factor depending on redshift and $\Gamma$ \citep{2001AJ....121.2879B}.

The {\it EP} will provide a great opportunity to investigate the LF and the ERD of the NSB-magnetar-driven XTs. It aims to monitor the XTs and variable objects in the soft X-ray band (0.5-4 keV) \citep{2015arXiv150607735Y,2017xru..conf..240Y}. A wide-field X-ray telescope (WXT) with a large FoV of $3600\;{\rm deg^{2}}$ and a narrow-field follow-up X-ray telescope (FXT) are on board {\it EP}. Using the novel lobster-eye optics, WXT may offer high sensitivity for X-ray all-sky monitors. Its sensitivity is $F_{\rm EP}\sim 5\times 10^{-11}\;{\rm erg\;cm^{-2}\;s^{-1}}$ corresponding to an exposure time of $10^3$ s. Considering the current observations, we adopt the $F_{\rm EP}$ as $1\times 10^{-9}\;{\rm erg\;cm^{-2}\;s^{-1}}$ and the derived $R(z)$ and LF to estimate the detectable XTs with {\it EP}/WXT. We consider a luminosity range of $\left[10^{44}, 10^{51}\right]\;{\rm erg\;s^{-1}}$ \citep{2014MNRAS.439.3916M,2017ApJ...835....7S,2018ApJ...857...95M}. Tab. \ref{tab:N} presents the estimated event rates of detectable XTs powered by the remnants for different EoSs. The type of remnant primarily affects the shape of the observed XT lightcurves. The HMNSs would rapidly evolve and collapse into a BH within $\sim 100$~ms. This short survival duration makes it challenging to detect the XTs they generate. Therefore, XTs detectable with {\it EP}/WXT are mainly driven by stable NSs or SMNSs, which can survive for an extended period as long as the NS is rotating rapidly. Once SMNSs lose significant angular momentum, they will collapse into a BH. The lightcurves of XTs produced by SMNSs typically exhibit a plateau followed by a very sharp drop, often referred to as an internal plateau. Our results indicate that $\sim 30\%$ XTs would show an internal plateau in the EoS DD2, while $\sim 98\%$ of the XTs generated by the EoSs AB-L and AB-N exhibit a plateau followed by a declining phase of approximately 1-2, known as the external plateau. In the EoS GM1 scenario, $94\%$ of lightcurves display an internal plateau. All lightcurves of XTs present a sharp drop in the EoS APR scenario. The total estimated detectable event rate is $\sim 1.8\times 10^{3}\;{\rm yr^{-1}}$ without considering any beaming effect. If the XT emission is collimated within a jet, the detection rate with {\it EP}/WXT would be significantly reduced. Considering a median jet opening angle of $\theta_{j} \sim 16^{\circ}$, as observed in sGRBs, a median beaming factor of $f_{\rm b}\approx 0.04$ can be calculated \citep{2015ApJ...815..102F}. Taking this into account, we estimate the detection event of XTs to be $\sim 70\;{\rm yr^{-1}}$ with {\it EP}/WXT. 

In our estimation, we adopt a broad luminosity range, extending the upper boundary to $10^{51}\;{\rm erg\;s^{-1}}$. This corresponds to the theoretical maximum power of a newborn magnetar central engine. However, XTs with such extreme luminosities are exceedingly rare in observations. As shown in the Fig.~\ref{fig:LF}, the brightest XT in our sample has a luminosity of only $\sim 2 \times 10^{49}\;{\rm erg\;s^{-1}}$. Therefore, lower the upper boundary to this value, i.e. $L\in[2\times 10^{44},2\times 10^{49}]\;{\rm erg\;s^{-1}}$, might provide a more realistic estimation of the detection rate. The reduced upper boundary is expected lower the detection rate since the $z_{\rm max}$ is also reduced. Apply the new luminosity range, we obtain a detection rate of $\sim 780\;{\rm yr^{-1}}$, half of the previous estimate. Accounting for the jet collimation effect, the rate further decreases to $\sim 31\;{\rm yr^{-1}}$, which is consistent with the {\it EP} observations during its first-year operation.

\begin{figure}
\center
\includegraphics[angle=0,scale=0.3]{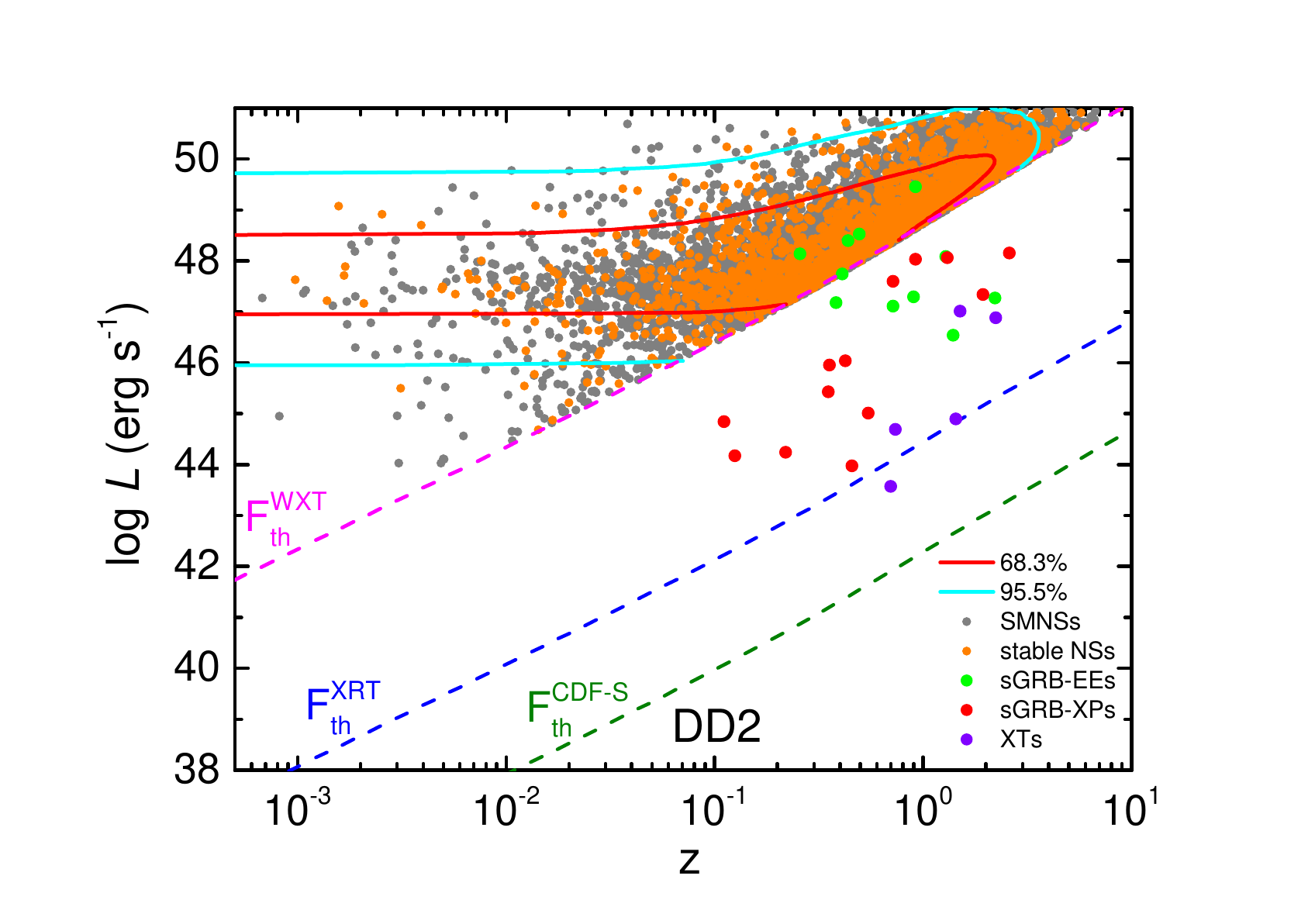}\\
\includegraphics[angle=0,scale=0.3]{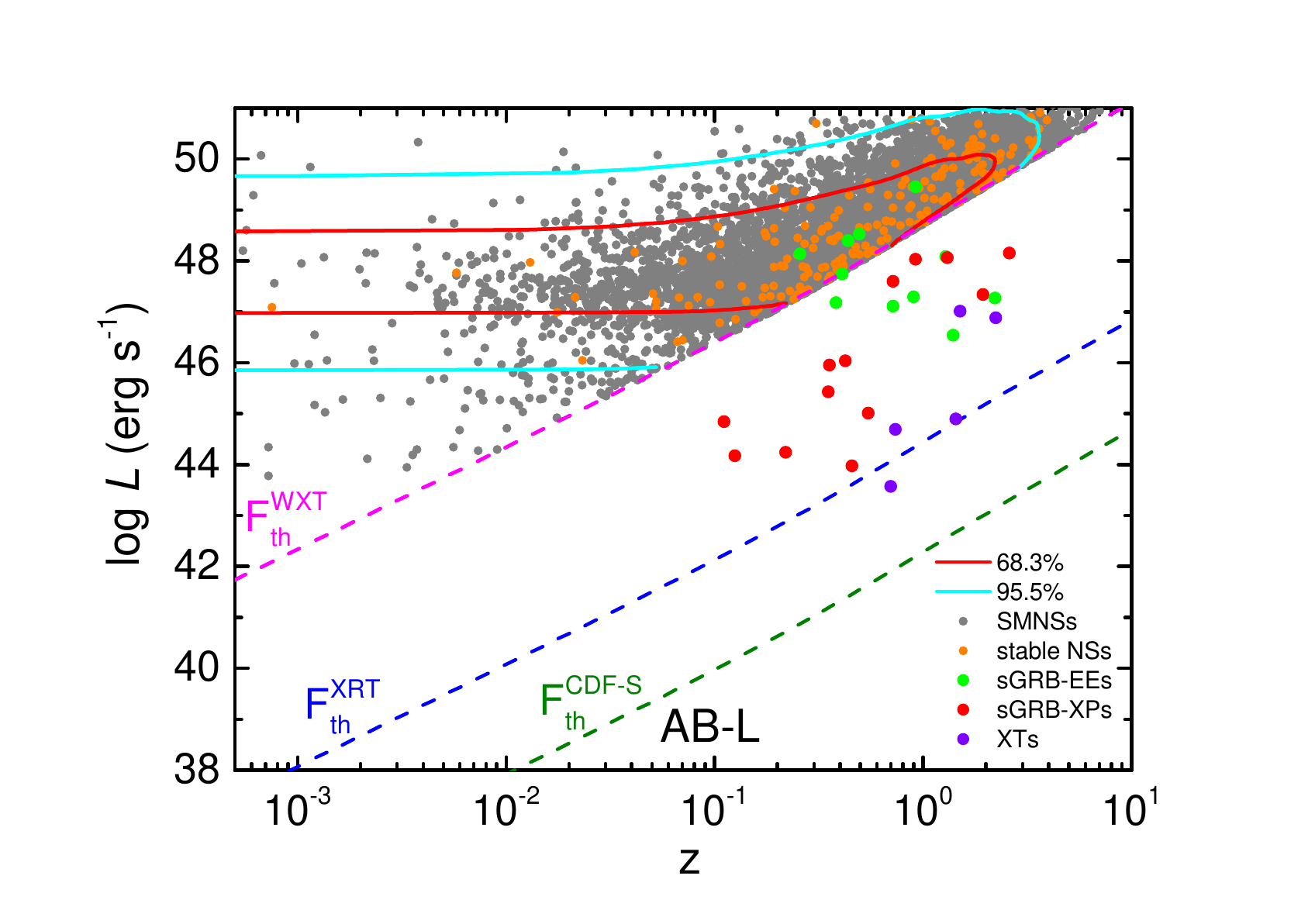}
\includegraphics[angle=0,scale=0.3]{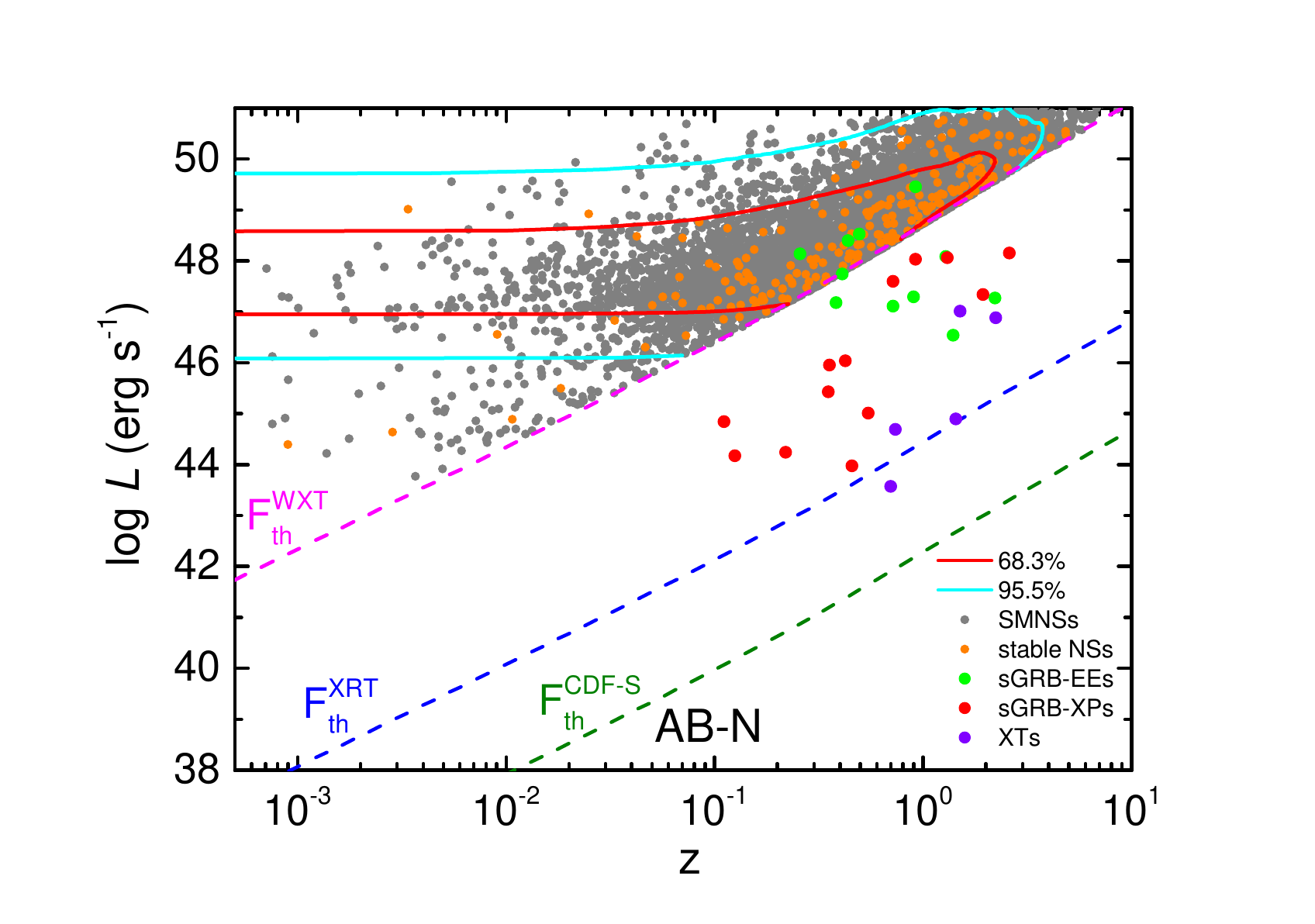}
\includegraphics[angle=0,scale=0.3]{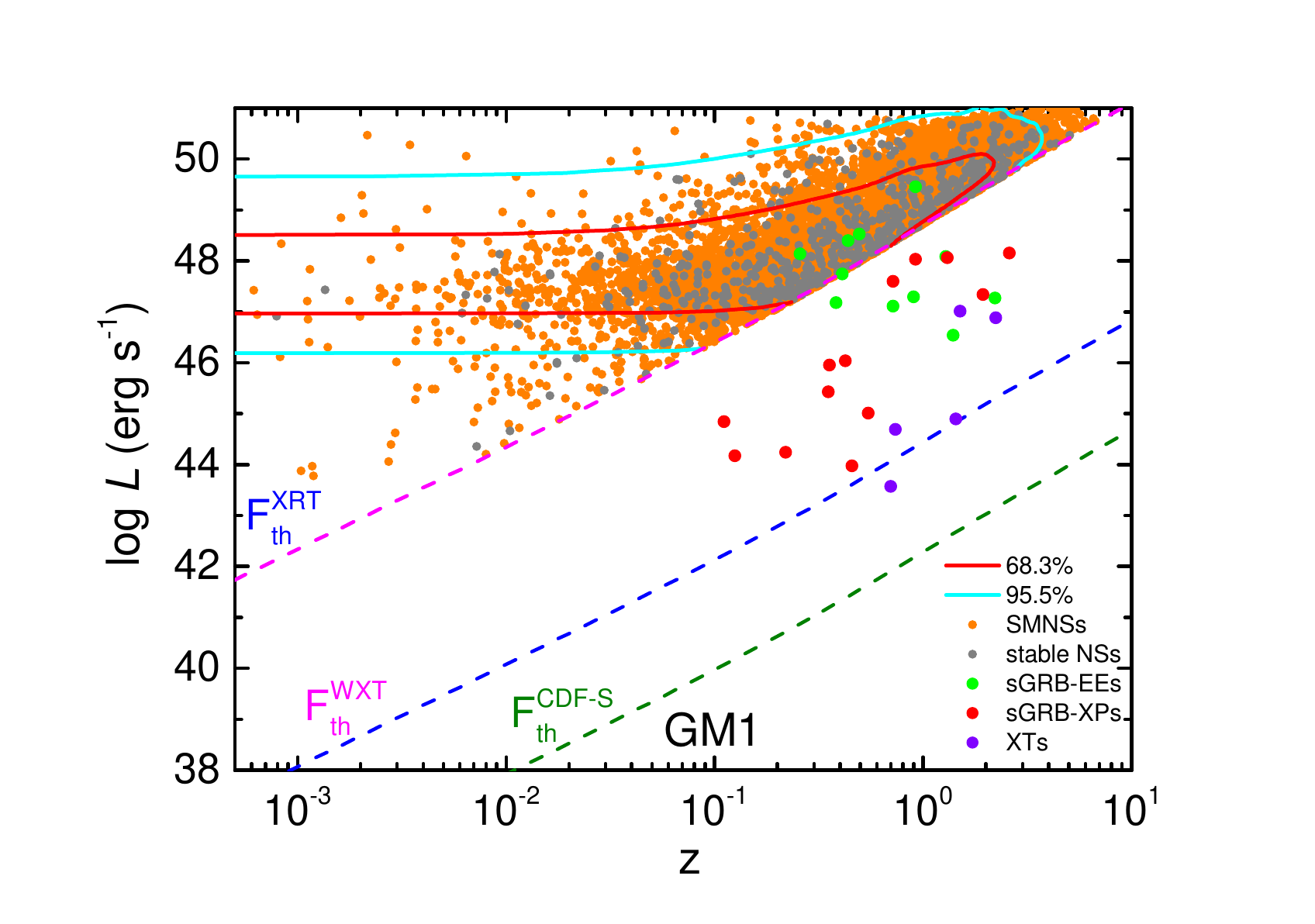}
\includegraphics[angle=0,scale=0.3]{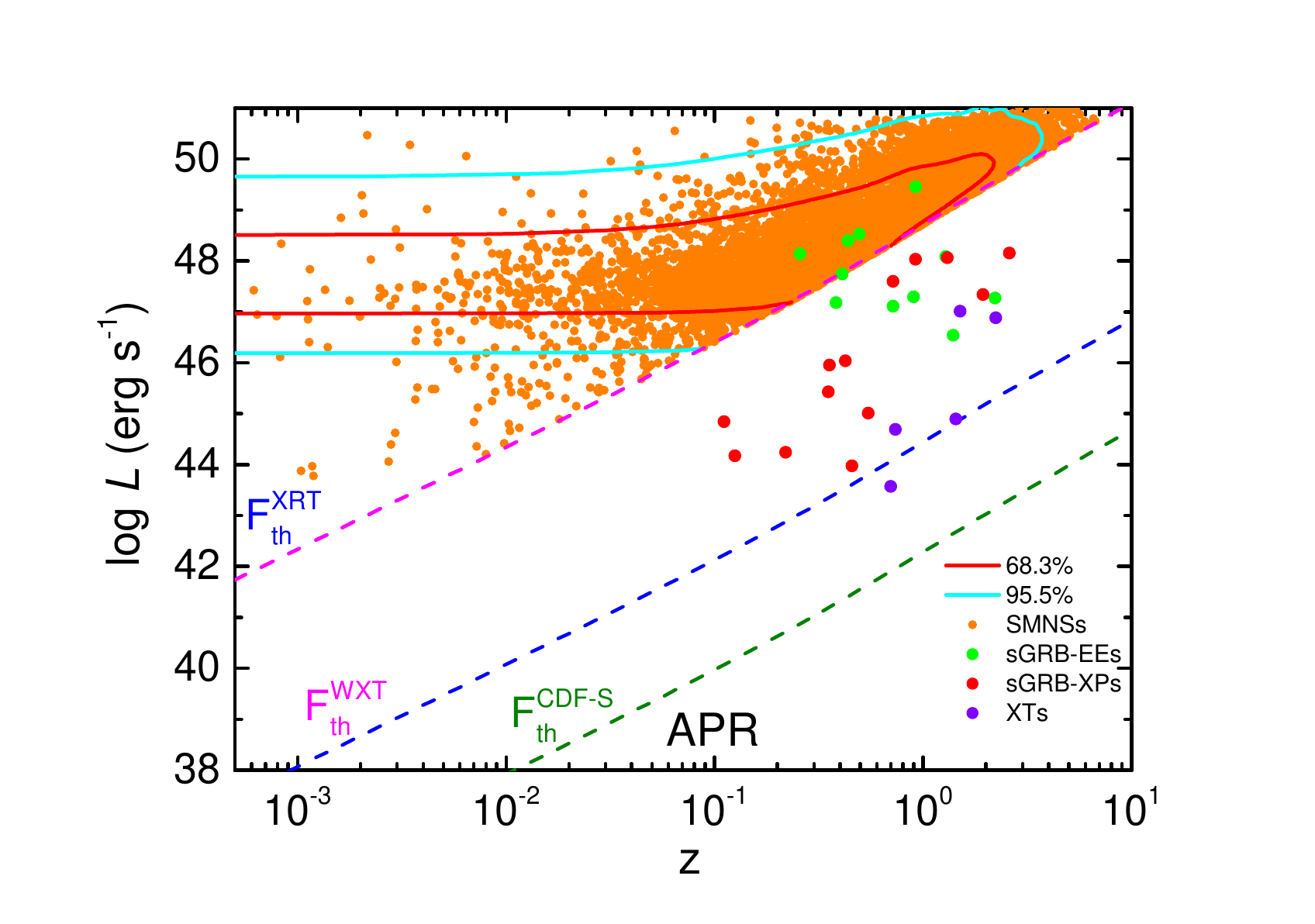}
\caption{Simulated probability distribution contours of observable XTs with the {\it EP}/WXT among the remnant types for different EoSs. The solid red, cyan, and black lines present the contours of $68.3\%$ and $95.5\%$ of the mock XTs distribution, respectively. The observed data (solid dots) are also shown. The dashed magenta, blue and olive lines stand for the roughly estimated sensitivity of {\it EP}/WXT ($1\times10^{-9}$ erg cm$^{-2}$ s$^{-1}$), $Swift$/XRT ($1\times10^{-13}$ erg cm$^{-2}$ s$^{-1}$), and {\it Chandra}/CDF-S ($4\times10^{-15}$ erg cm$^{-2}$ s$^{-1}$), respectively. The data and the sensitivities of $Swift$/XRT and {\it Chandra}/CDF-S have been converted to the 0.5-4~keV band.}
\label{fig:dis}
\end{figure}

We perform Monte Carlo simulations to create a mock sample of NSB-magnetar-driven XTs for visualizing the probability distribution of detectable such XTs with the {\it EP}/WXT. We generate $2\times 10^{5}$ XTs from different remnant types corresponding to different EoSs. Each XT in this sample is characterized by a set of parameters $(z_{\rm sim}, L_{\rm sim}, \Gamma_{\rm sim})$, randomly chosen from the normalized probabilities distribution of the $R(z)$ where $z\in[0, 10]$, $\Phi(L)$ where $L_{\rm iso}\in [10^{44},10^{51}]$ ${\rm erg\; \;\rm s^{-1}}$ and the Gaussian distribution of $\Gamma$ as presented in \citep{2021MNRAS.508.2505Z}. We then screen the generated XTs using the WXT threshold\footnote{ The designed trigger threshold of the {\it EP}/WXT is 10 mCrab ($\sim 5\times 10^{-10}\;{\rm erg}\;{\rm cm^{-2}}\;{\rm s^{-1}}$ assuming a photon index of 2.1 in 0.5-4 keV band) with 600 second exposure time; https://www.nssdc.ac.cn/). Most XTs triggered by WXT in 2024 are brighter than $10^{-9}\;{\rm erg\;cm^{-2}\;s^{-1}}$. Thus, we adopt a robust threshold of $10^{-9}\;{\rm erg\;cm^{-2}\;s^{-1}}$ in our analysis.} $F_{\rm sim}\geq F_{\rm EP}=1\times 10^{-9}\;{\rm erg\;cm^{-2}\;s^{-1}}$, where $F_{\rm sim} = L_{\rm sim} / 4 \pi D^2_{\rm L}(z_{\rm sim}) k(z_{\rm sim}, \Gamma_{\rm sim})$. Fig. \ref{fig:dis} illustrates the distributions and the density contours of the mock XT sample among the remnant types for different EoSs in the $\log (1 + z)-\log L$ plane. For the EoS DD2, we obtain 2089 detectable XTs with an internal plateau and 5968 detectable XTs with an external plateau, respectively. In the case of the EoSs AB-L and AB-N, we find 196 detectable XTs with an internal plateau and 7840 detectable XTs with an external plateau, and 221 detectable XTs with an internal plateau and 7898 detectable XTs with an external plateau, respectively. For the EoS GM1, we find 7336 detectable XTs with an internal plateau and 611 detectable XTs with an external plateau.  Finally, for the EoS APR, we obtain 8028 detectable XTs with an internal plateau. In summary, we estimate that $\sim 8\times 10^3$ XTs would be detectable with {\it EP}/WXT out of the $2\times 10^{5}$ XTs generated, which is comparable to the distribution of the XTs detected by {\it EP}/WXT over three years of mission operation.

We also compare the XTs listed in Tab. \ref{tab:paras} with the detection thresholds of {\it Chandra}, {\it EP}/WXT, and plot them in Fig. \ref{fig:dis} for comparison. It should be noted that all data points and threshold lines in this figure are rescaled to the {\it EP}/WXT energy band of 0.5-4 keV. From the plot, we can observe that the sGRB-EEs represent the high-luminosity end of the XT population and about half of the XTs extracted from the sGRB-EEs would be detectable with {\it EP}/WXT. However, {\it EP}/WXT may not be able to capture XTs similar to those found in the X-ray afterglow plateau data and the CDS-S data. Nevertheless, {\it EP}/WXT still possesses an excellent capability to establish a significant sample of these types of XTs, which could enrich our understanding of this XT population.

\begin{figure}
\center
\includegraphics[angle=0,scale=0.3]{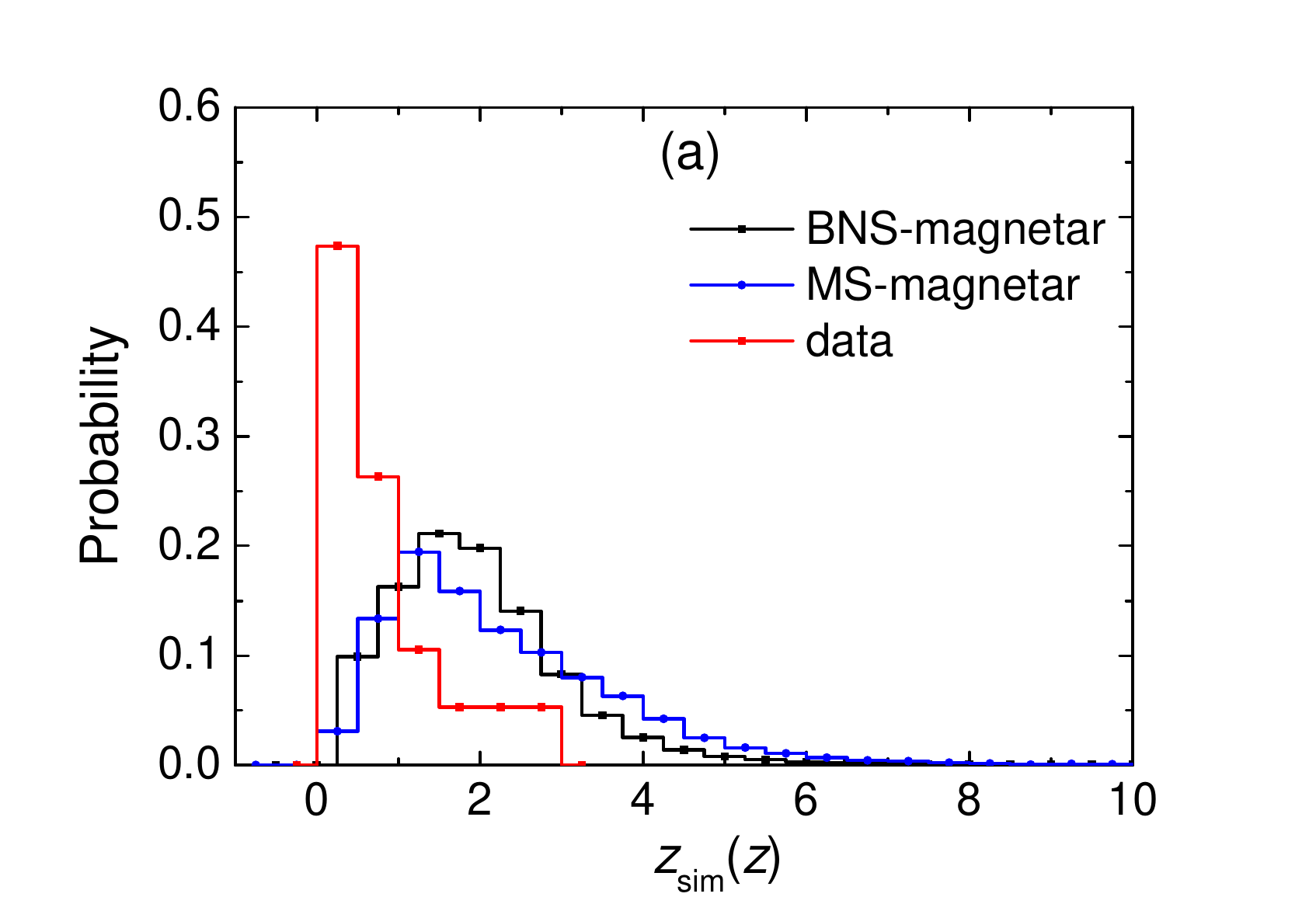}
\includegraphics[angle=0,scale=0.3]{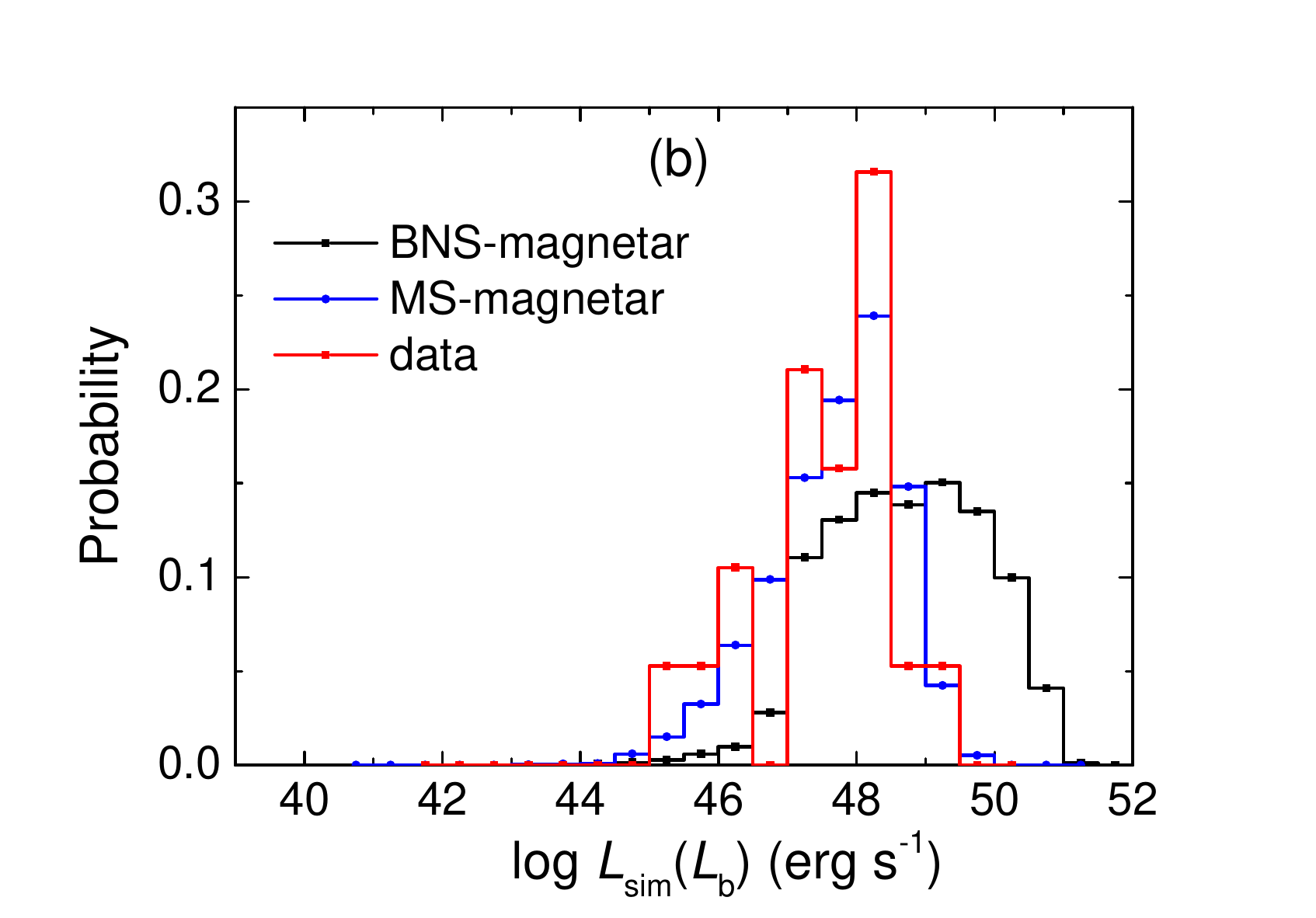}
\caption{Distributions of $z_{\rm sim}$ and $L_{\rm sim}$ for the mock sample for NSB-magnetars (black lines) and MS-magnetars (blue lines), respectively. And the $z$ and $L_{\rm b}$ distributions of our sample.}
\label{fig:zL}
\end{figure}

Without considering the specific light curve shapes of XTs, we examine the distributions of $z_{\rm sim}$ and $L_{\rm sim}$ in our mock sample, using the results obtained with EoS DD2 as an example. The distributions are shown in Fig. \ref{fig:zL}. In the same figure, we also compare the distributions of $z$ and $L_{\rm b}$ in our sample with those of the XTs driven by MS-magnetars taken from \cite{2020ApJ...894...52X}. We find that the distribution of $z_{\rm sim}$ in our mock sample is roughly consistent with that taken from \cite{2020ApJ...894...52X}. The distribution of $L_{\rm b}$ is broad, and the higher luminosity end aligns well with the distribution of $L_{\rm sim}$ for XTs driven by MS-magnetars. This is mainly due to the high-luminosity XTs in our sample corresponding to the extreme events (EEs) observed in sGRBs. According to the distributions in Fig. \ref{fig:zL}, we find that there is no way to tell from $z$ and $L$ alone whether the XTs driven by an MS-magnetar or an NSB-magnetar. Additionally, it is intriguing to note that some nearby XTs with $z<0.01$ could potentially be discovered in the {\it EP}/WXT survey.

\section{Discussions}
\label{sec:Dis}

\subsection{Uncertainties in deriving the detection rate}
As shown in Eq.~\ref{eq:dNdL}, our detection rate estimation relies on several assumptions, which are introducing potential uncertainties. First, we assume that every newborn magnetar successfully produces a XT, as we are unable to acquire the failure rate. This assumption may lead to an overestimation of the detection rate. Second, we impose a luminosity range of $[L_{\rm min}, L_{\rm max}]$ for NSB-magnetar-driven XTs. So that the choices of $L_{\rm min}$ and $L_{\rm max}$ are critical in deriving the detection rate. A larger $L_{\rm max}$ is expected to increase the detection rate, as more XTs surpass the instrument threshold flux, whereas a lower $L_{\rm min}$ reduces the detection rate by increasing the number of XTs that fall below the threshold. Finally, additional factors may prevent some XTs from being detected, such as the jet collimation effect mentioned earlier and various absorption effects in the propagation X-ray emission. Additionally, as emphasized in Sec.~\ref{sec:LF}, the derived LF for NSB-magnetar-driven XTs is induced with uncertainties from instrument selection effect and small sample size. 

Comparison of our results with those from previous studies may provide more valuable insights. \cite{2022A&A...663A.168Q} conduct a comprehensive investigation on {\it Chandra}-detected XTs, indiscriminating their physical origin. Their work adopts a different approach from ours to estimate the detection rate. By extrapolating the {\it Chandra} detection of XTs with instrument parameters like the effective area and the exposure time accounted, they estimate that the {\it EP} detection rate of the XTs are $\sim 35-69\;{\rm yr^{-1}}$. Adopting the same limit flux\footnote{$F_{{\rm lim}, {\it EP}{\rm /WXT}} \approx 5\times10^{-10}\;{\rm erg}\;{\rm cm^{-2}}\;{\rm s^{-1}}$. This limit flux was approximated as 10 times the WXT threshold flux $F_{{\rm th}, {\it EP}{\rm /WXT}} = 5 \times 10^{-11}\;{\rm erg\;cm^{-2}\;s^{-1}}$ at a $10^{3}\;{\rm s}$ exposure, to avoid the influence of the Poisson noise effects \citep{2022A&A...663A.168Q}.} of the {\it EP}/WXT used in their estimation, we estimate that the detection rate is $\sim 1638\;{\rm yr^{-1}}$ without considering the jet collimation effect, which is significantly higher than theirs. There are several expectable reasons for the discrepancy. The first is that our method tends to overestimate the detection rate, as already discussed. If taking the jet collimation factor $f_{\rm b} \approx 0.04$ into consideration, our detection rate is reduced to $\sim 66\;{\rm yr^{-1}}$ which becomes comparable to their prediction. The second is that their prediction depends solely on the number of {\it Chandra}-detected XTs and instrument parameters and ignores the physical origins of the XTs so that their results would also be influenced by the selection effect. Considering that we are focusing on a specific subclass of XTs, our detection rate remains slightly higher after incorporating the jet collimation effect. This discrepancy may indicate that not every newborn magnetar can produce an XT.

\subsection{EoSs and the type of XTs}
The EoSs of NSs remain an open question in astrophysics, and numerous studies have been conducted to constrain the EoS using various observational data. For example, \cite{2014PhRvD..89d7302L} analyzed eight individual sGRBs and compared the observed data with five different EoSs. Their findings suggested that the EoS GM1 was most consistent with the observed data. In a more extensive analysis, \cite{2016PhRvD..93d4065G} employed Monte Carlo simulations to test these five EoSs against observational constraints from a broader sGRB sample. They concluded that only the EoS GM1 could satisfy the constraints on the fraction of supra-massive neutron stars (NSs), while the other four EoSs could not. Further investigations by \cite{2016PhRvD..94h3010L} expanded the analysis to include additional NS EoSs and quark star (QS) EoSs. They found that all except one (EoS BCPM) could satisfy the observational constraints on supra-massive NSs/QSs. Regarding specific observations, \cite{2021ApJ...913...27L} and \cite{2021RAA....21...47L} proposed that the EoS DD2 is favored by the observations of GW 170817 and CDF-S XT2. However, it should be noted that there are different possibilities for the remnant of GW 170817, and this may lead to very different constraints of NS EoSs. The population synthesis results could provide a novel and valuable method for constraining the EoS of NSs.

Considering the remnant types for different EoSs, our population synthesis results indicate that $\sim 30\%$ of detectable XTs with an internal plateau are expected for the EoS DD2. For the EoSs AB-L and AB-N, $\sim 98\%$ of the detectable XTs with an external plateau, while $\sim 94\%$ of the detectable XTs in the EoS GM1 and all detectable XTs in the EoS APR are expected to display an internal plateau. In a systematic analysis conducted by \cite{2021MNRAS.508.2505Z}, they investigated the magnetar model using sGRB X-ray afterglow data observed by {\it Swift}/XRT. Combining their results with our sample, we obtained 29 sGRBs with XPs, in which 8 sGRBs presented an internal plateau. Assuming these XPs are the candidate XTs powered by NSs, it suggests that about $28\%$ of the XPs observed by {\it Swift}/XRT show an internal plateau, consistent with the results obtained for the EoS DD2. These results may indicate that the EoS DD2 could be more satisfied with the observations without considering the selection effect of the instrument. However, it is important to note that the bursts analyzed in our study represent only a small fraction of the observations with {\it Swift}/XRT. To further study the EoSs of NSs and derive more robust conclusions, a larger sample of XTs is needed for comprehensive analysis.

\subsection{$N$ as a function of $F_{\rm th}$}
\begin{figure}
\center
\includegraphics[angle=0,scale=0.50]{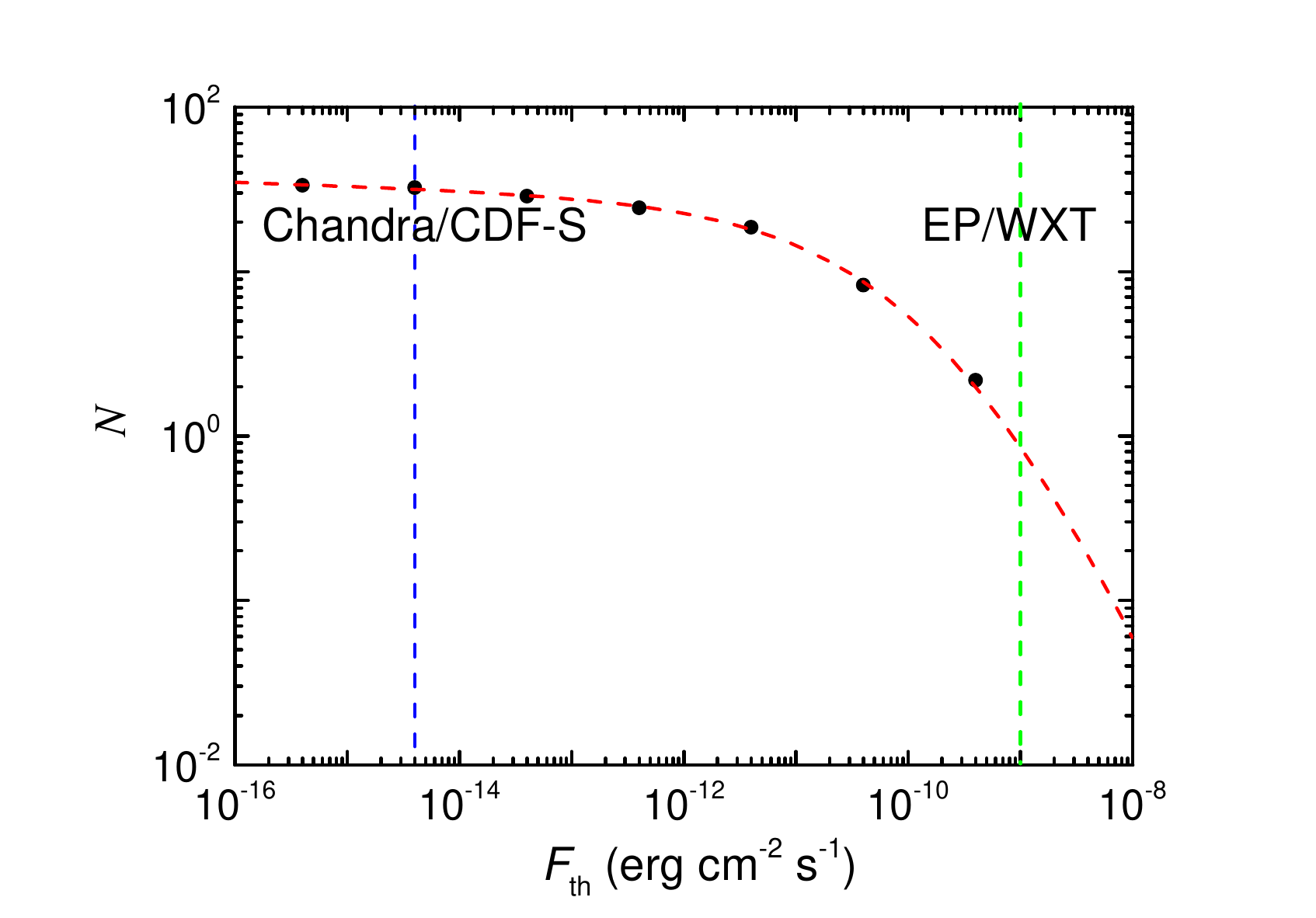}
\caption{Observable event number as a function of instrument flux threshold based on the cosmic event rate density and the LF derived from our analysis. The red dashed line presents the best fit. The blue and green dashed lines mark the sensitivities for {\it Chandra}/CDF-S and {\it EP}/WXT, respectively.}
\label{fig:N}
\end{figure}

To further explore the differences in detectable event rates, we examine the relationship between the observable event number $N$ and the instrumental threshold $F_{\rm th}$. In Fig. \ref{fig:N}, we illustrate $N$ as a function of $F_{\rm th}$, incorporating the FoV of {\it Chandra} $(\Omega_{\rm cha}$), along with the derived $R(z)$ and the LF. One can observe that the relationship between $N$ and $F_{\rm th}$ is well described by a broken power-law function, with best-fit parameters $\alpha_{1}=0.02\pm 0.01$, $\alpha_{2}=1.55\pm 3.55$, and a break at $F_{\rm th,b}=(3.00\pm 0.25)\times 10^{-9}\;{\rm erg\;cm^{-2}\;s^{-1}}$. Based on this result, we find that the expected event rate of XTs observed by {\it Chandra} ($N_{\rm Cha}$) is about one order of magnitude larger than $N_{\rm EP}$ within the same energy band and FoV, owing to the lower $F_{\rm th}$ of {\it Chandra} compared to {\it EP}/WXT. However, when considering the FoVs of both instruments and defining the ratio ($\Re$) of the event rate observable with the {\it EP}/WXT to that with {\it Chandra} as $\Re=(N_{\rm EP}\Omega_{\rm EP})/(N_{\rm Cha}\Omega_{\rm Cha})$, we derive $\Re \simeq 50$ in the 0.5-7 keV band, which indicates that the number of events detectable by {\it EP}/WXT is approximately 50 times larger than the detected by {\it Chandra}, primarily due to the significantly wider FoV {\it EP}/WXT.

\section{Conclusions}
\label{sec:Con}

Based on the results of the population synthesis numerical simulation, we have investigated the event rate densities $R(z)$ of NSB mergers as a function of $z$ for different EoSs. By considering the M33.A model simulation results and assuming that a newborn magnetar is the outcome of an NSB event with a total mass $M_{\rm tot}\leq 1.3 M_{\rm TOV}$, we estimate the global local event rate density of NSB-magnetars to be $R_{0}\sim300\;{\rm Gpc^{-3}}\;{\rm yr^{-1}}$. The cosmic history of $R(z)$ peaks at $z_{\rm peak} = 1.81$ and exhibits a rapid decrease as $R\propto z^{-3.85}$ for $z>4$. Regardless of the instrumental selection effects and based on the plausible XTs observed in the {\it Chandra}/CDF-S data and the early X-ray afterglow data of sGRBs, we have parameterized the LF of NSB-magnetar XTs as a combination of a single power-law function and a broken power-law function with indices of $a=-1.03$, $\beta_{1}=-0.28$, and $\beta_{2}=-1.66$, as well as a break luminosity $L_{\rm b}=4.38\times 10^{47}\;{\rm erg}\;{\rm s^{-1}}$. Our analysis predicts that the detectable rate of the XTs with {\it EP}/WXT is $\sim 780\;{\rm yr^{-1}}$ in the luminosity range of $[2\times 10^{44},2\times 10^{49}]\;{\rm erg}\; \; {\rm s^{-1}}$ without considering any beaming effect. When considering an ejecta opening angle of $16^{\circ}$, similar to that observed in sGRBs, this number significantly drops to $\sim 31$ $\rm yr^{-1}$. According to the results reported by the {\it EP}/WXT in the past year, our result considering the jet opening angle does not violate the current observations. However, it is noted that our estimation is subject to uncertainties arising from the LF derivation, due to the instrumental selection effect and the limited sample size. Currently, we are unable to quantify their impact on the estimated detection rate. We expect that the {\it EP} observations will be crucial in reducing these uncertainties. Furthermore, according to the population synthesis results, we can also determine the fraction of the types of XTs among the remnants for different EoSs. Our results suggest that the EoS DD2 more closely aligns with observations when not considering the selection effect of the instrument.

These XTs as the NSB post-merger products are the most prospective observable EM counterparts of the GW sources. Assuming the mass of the NSB merger is below $2.5M_{\odot}$, \cite{2020A&A...636A.104B} gives that the local merger rate density is $R_{0}=321$ $\rm Gpc^{-3}$ $\rm yr^{-1}$, the GW detection rate for LIGO/Virgo's mid-high sensitivity curve is $R_{\rm GW}=1.329$ yr$^{-1}$, with the capability to detect sources up to a redshift of $z_{\rm dis}=0.082$. These findings highlight that, despite a high NSB merger rate, current instruments are likely to detect only a limited number of GW events. When adopting $R_{0}=321$ $\rm Gpc^{-3}$ $\rm yr^{-1}$, the detectable event rate for NSB-magnetar XTs is $3.80$ yr$^{-1}$ at $z_{\rm dis}$. If considering $R_{0}= 300$ $\rm Gpc^{-3}$ $\rm yr^{-1}$, the detectable event rate is $3.53$ yr$^{-1}$.

\cite{2017ApJ...835....7S} investigated the peak LF and event rate density of the XTs powered by the NSB-magnetars associated with sGRBs via Monte Carlo simulations. They proposed a bimodal LF, peaking at $2.5\times 10^{46}$ erg s$^{-1}$ and $4\times 10^{49}$ erg s$^{-1}$, corresponding to light-of-sight to the trapped (the direction where X-rays are initially trapped by the dynamical ejecta) and the free (the direction where no sGRB is observed) zones, respectively. They inferred an event rate density of a few tens Gpc$^{-3}$ yr$^{-1}$  at $L>10^{45}$ erg s$^{-1}$. Note that the GRB jet emission could be choked by extended material and stalled far below the photosphere \citep{2016PhRvD..93h3003S} or the misalignment of their jet axis to the light of sight \citep{2019ApJ...877..153Z,2021ApJ...921L...1Z}, the rate estimated by \cite{2017ApJ...835....7S} should be only a rough lower limit of the XT rate.

\begin{acknowledgments}
We very appreciate helpful comments and suggestions from the referee.
We acknowledge the use of the public data from the {\em Swift} data archive and the UK {\em Swift} Science Data Center. This work is supported by the National Key R\&D Project (No. 2024YFA1611700) and the National Natural Science Foundation of China (Grant No. 12403056 and 12133003).
\end{acknowledgments}


\bibliographystyle{aasjournal}

\begin{thebibliography}{}
\expandafter\ifx\csname natexlab\endcsname\relax\def\natexlab#1{#1}\fi
\providecommand{\url}[1]{\href{#1}{#1}}
\providecommand{\dodoi}[1]{doi:~\href{http://doi.org/#1}{\nolinkurl{#1}}}
\providecommand{\doeprint}[1]{\href{http://ascl.net/#1}{\nolinkurl{http://ascl.net/#1}}}
\providecommand{\doarXiv}[1]{\href{https://arxiv.org/abs/#1}{\nolinkurl{https://arxiv.org/abs/#1}}}

\bibitem[Abbott et al.(2017b)]{2017ApJ...848L..12A} Abbott, B.~P., Abbott, R., Abbott, T.~D., et al.\ 2017, \apjl, 848, L12. doi:10.3847/2041-8213/aa91c9
\bibitem[Abbott et al.(2019)]{2019PhRvX...9c1040A} Abbott, B.~P., Abbott, R., Abbott, T.~D., et al.\ 2019, Physical Review X, 9, 031040. doi:10.1103/PhysRevX.9.031040
\bibitem[Abbott et al.(2017a)]{2017PhRvL.119p1101A} Abbott, B.~P., Abbott, R., Abbott, T.~D., et al.\ 2017, \prl, 119, 161101. doi:10.1103/PhysRevLett.119.161101
\bibitem[Akmal et al.(1998)]{1998PhRvC..58.1804A} Akmal, A., Pandharipande, V.~R., \& Ravenhall, D.~G.\ 1998, \prc, 58, 1804. doi:10.1103/PhysRevC.58.1804
\bibitem[Anand et al.(2018)]{2018MNRAS.481.4332A} Anand, N., Shahid, M., \& Resmi, L.\ 2018, \mnras, 481, 4332. doi:10.1093/mnras/sty2530
\bibitem[Arnett \& Bowers(1977)]{1977ApJS...33..415A} Arnett, W.~D. \& Bowers, R.~L.\ 1977, \apjs, 33, 415. doi:10.1086/190434
\bibitem[Bauer et al.(2017)]{2017MNRAS.467.4841B} Bauer, F.~E., Treister, E., Schawinski, K., et al.\ 2017, \mnras, 467, 4841. doi:10.1093/mnras/stx417
\bibitem[Baumgarte et al.(2000)]{2000ApJ...528L..29B} Baumgarte, T.~W., Shapiro, S.~L., \& Shibata, M.\ 2000, \apjl, 528, L29. doi:10.1086/312425
\bibitem[Belczynski et al.(2020)]{2020A&A...636A.104B} Belczynski, K., Klencki, J., Fields, C.~E., et al.\ 2020, \aap, 636, A104. doi:10.1051/0004-6361/201936528
\bibitem[Belczynski et al.(2002)]{2002ApJ...572..407B} Belczynski, K., Kalogera, V., \& Bulik, T.\ 2002, \apj, 572, 407. doi:10.1086/340304
\bibitem[Belczynski et al.(2008)]{2008ApJS..174..223B} Belczynski, K., Kalogera, V., Rasio, F.~A., et al.\ 2008, \apjs, 174, 223. doi:10.1086/521026
\bibitem[Bloom et al.(2001)]{2001AJ....121.2879B} Bloom, J.~S., Frail, D.~A., \& Sari, R.\ 2001, \aj, 121, 2879. doi:10.1086/321093
\bibitem[Chu et al.(2022)]{2022MNRAS.509.1557C} Chu, Q., Yu, S., \& Lu, Y.\ 2022, \mnras, 509, 1557. doi:10.1093/mnras/stab2882
\bibitem[Cook et al.(1994)]{1994ApJ...424..823C} Cook, G.~B., Shapiro, S.~L., \& Teukolsky, S.~A.\ 1994, \apj, 424, 823. doi:10.1086/173934
\bibitem[Dai \& Lu(1998)]{1998A&A...333L..87D} Dai, Z.~G. \& Lu, T.\ 1998, \aap, 333, L87. doi:10.48550/arXiv.astro-ph/9810402
\bibitem[Duncan \& Thompson(1992)]{1992ApJ...392L...9D} Duncan, R.~C. \& Thompson, C.\ 1992, \apjl, 392, L9. doi:10.1086/186413
\bibitem[Evans et al.(2010)]{2010A&A...519A.102E} Evans, P.~A., Willingale, R., Osborne, J.~P., et al.\ 2010, \aap, 519, A102. doi:10.1051/0004-6361/201014819
\bibitem[Fong et al.(2015)]{2015ApJ...815..102F} Fong, W., Berger, E., Margutti, R., et al.\ 2015, \apj, 815, 102. doi:10.1088/0004-637X/815/2/102
\bibitem[Fryer et al.(2012)]{2012ApJ...749...91F} Fryer, C.~L., Belczynski, K., Wiktorowicz, G., et al.\ 2012, \apj, 749, 91. doi:10.1088/0004-637X/749/1/91
\bibitem[Gao et al.(2020)]{2020FrPhy..1524603G} Gao, H., Ai, S.-K., Cao, Z.-J., et al.\ 2020, Frontiers of Physics, 15, 24603. doi:10.1007/s11467-019-0945-9
\bibitem[Gao et al.(2016)]{2016PhRvD..93d4065G} Gao, H., Zhang, B., \& L{\"u}, H.-J.\ 2016, \prd, 93, 044065. doi:10.1103/PhysRevD.93.044065
\bibitem[Gehrels(1986)]{1986ApJ...303..336G} Gehrels, N.\ 1986, \apj, 303, 336. doi:10.1086/164079
\bibitem[Giacomazzo et al.(2015)]{2015ApJ...809...39G} Giacomazzo, B., Zrake, J., Duffell, P.~C., et al.\ 2015, \apj, 809, 39. doi:10.1088/0004-637X/809/1/39
\bibitem[Giacomazzo \& Perna(2013)]{2013ApJ...771L..26G} Giacomazzo, B. \& Perna, R.\ 2013, \apjl, 771, L26. doi:10.1088/2041-8205/771/2/L26
\bibitem[Glendenning \& Moszkowski(1991)]{1991PhRvL..67.2414G} Glendenning, N.~K. \& Moszkowski, S.~A.\ 1991, \prl, 67, 2414. doi:10.1103/PhysRevLett.67.2414
\bibitem[Hotokezaka et al.(2013)]{2013PhRvD..87b4001H} Hotokezaka, K., Kiuchi, K., Kyutoku, K., et al.\ 2013, \prd, 87, 024001. doi:10.1103/PhysRevD.87.024001
\bibitem[Hu et al.(2024)]{2024GCN.36053....1H} Hu, J.~W., Zhao, D.~H., Liu, Y., et al.\ 2024, GRB Coordinates Network, Circular Service, No. 36053, 36053
\bibitem[Kumar \& Zhang(2015)]{2015PhR...561....1K} Kumar, P. \& Zhang, B.\ 2015, \physrep, 561, 1. doi:10.1016/j.physrep.2014.09.008
\bibitem[Lasky et al.(2014)]{2014PhRvD..89d7302L} Lasky, P.~D., Haskell, B., Ravi, V., et al.\ 2014, \prd, 89, 047302. doi:10.1103/PhysRevD.89.047302
\bibitem[Li et al.(2021)]{2021ApJ...913...27L} Li, A., Miao, Z., Han, S., et al.\ 2021, \apj, 913, 27. doi:10.3847/1538-4357/abf355
\bibitem[Li et al.(2016)]{2016PhRvD..94h3010L} Li, A., Zhang, B., Zhang, N.-B., et al.\ 2016, \prd, 94, 083010. doi:10.1103/PhysRevD.94.083010
\bibitem[Li et al.(2021)]{2021ApJS..252...16L} Li, X.~J., Zhang, Z.~B., Zhang, X.~L., et al.\ 2021, \apjs, 252, 16. doi:10.3847/1538-4365/abd3fd
\bibitem[Lian et al.(2024)]{2024GCN.36091....1L} Lian, T.~Y., Pan, X., Ling, Z.~X., et al.\ 2024, GRB Coordinates Network, Circular Service, No. 36091, 36091
\bibitem[Lian et al.(2024)]{2024GCN.36086....1L} Lian, T.~Y., Pan, X., Ling, Z.~X., et al.\ 2024, GRB Coordinates Network, Circular Service, No. 36086, 36086
\bibitem[Lien et al.(2014)]{2014ApJ...783...24L} Lien, A., Sakamoto, T., Gehrels, N., et al.\ 2014, \apj, 783, 24. doi:10.1088/0004-637X/783/1/24
\bibitem[Lyons et al.(2010)]{2010MNRAS.402..705L} Lyons, N., O'Brien, P.~T., Zhang, B., et al.\ 2010, \mnras, 402, 705. doi:10.1111/j.1365-2966.2009.15538.x
\bibitem[L{\"u} et al.(2021)]{2021RAA....21...47L} L{\"u}, H.-J., Yuan, Y., Lan, L., et al.\ 2021, Research in Astronomy and Astrophysics, 21, 047. doi:10.1088/1674-4527/21/2/47
\bibitem[L{\"u} \& Zhang(2014)]{2014ApJ...785...74L} L{\"u}, H.-J. \& Zhang, B.\ 2014, \apj, 785, 74. doi:10.1088/0004-637X/785/1/74
\bibitem[L{\"u} et al.(2018)]{2018MNRAS.480.4402L} L{\"u}, H.-J., Zou, L., Lan, L., et al.\ 2018, \mnras, 480, 4402. doi:10.1093/mnras/sty2176
\bibitem[L{\"u} et al.(2015)]{2015ApJ...805...89L} L{\"u}, H.-J., Zhang, B., Lei, W.-H., et al.\ 2015, \apj, 805, 89. doi:10.1088/0004-637X/805/2/89
\bibitem[Madau \& Fragos(2017)]{2017ApJ...840...39M} Madau, P. \& Fragos, T.\ 2017, \apj, 840, 39. doi:10.3847/1538-4357/aa6af9
\bibitem[Margalit \& Metzger(2017)]{2017ApJ...850L..19M} Margalit, B. \& Metzger, B.~D.\ 2017, \apjl, 850, L19. doi:10.3847/2041-8213/aa991c
\bibitem[Metzger et al.(2011)]{2011MNRAS.413.2031M} Metzger, B.~D., Giannios, D., Thompson, T.~A., et al.\ 2011, \mnras, 413, 2031. doi:10.1111/j.1365-2966.2011.18280.x
\bibitem[Metzger \& Piro(2014)]{2014MNRAS.439.3916M} Metzger, B.~D. \& Piro, A.~L.\ 2014, \mnras, 439, 3916. doi:10.1093/mnras/stu247
\bibitem[Metzger et al.(2018)]{2018ApJ...857...95M} Metzger, B.~D., Beniamini, P., \& Giannios, D.\ 2018, \apj, 857, 95. doi:10.3847/1538-4357/aab70c
\bibitem[Quirola-V{\'a}squez et al.(2022)]{2022A&A...663A.168Q} Quirola-V{\'a}squez, J., Bauer, F.~E., Jonker, P.~G., et al.\ 2022, \aap, 663, A168. doi:10.1051/0004-6361/202243047
\bibitem[Quirola-V{\'a}squez et al.(2023)]{2023A&A...675A..44Q} Quirola-V{\'a}squez, J., Bauer, F.~E., Jonker, P.~G., et al.\ 2023, \aap, 675, A44. doi:10.1051/0004-6361/202345912
\bibitem[Rezzolla et al.(2011)]{2011ApJ...732L...6R} Rezzolla, L., Giacomazzo, B., Baiotti, L., et al.\ 2011, \apjl, 732, L6. doi:10.1088/2041-8205/732/1/L6
\bibitem[Rosswog et al.(2013)]{2013MNRAS.430.2585R} Rosswog, S., Piran, T., \& Nakar, E.\ 2013, \mnras, 430, 2585. doi:10.1093/mnras/sts708
\bibitem[Rowlinson et al.(2010)]{2010MNRAS.409..531R} Rowlinson, A., O'Brien, P.~T., Tanvir, N.~R., et al.\ 2010, \mnras, 409, 531. doi:10.1111/j.1365-2966.2010.17354.x
\bibitem[Rowlinson et al.(2013)]{2013MNRAS.430.1061R} Rowlinson, A., O'Brien, P.~T., Metzger, B.~D., et al.\ 2013, \mnras, 430, 1061. doi:10.1093/mnras/sts683
\bibitem[Senno et al.(2016)]{2016PhRvD..93h3003S} Senno, N., Murase, K., \& M{\'e}sz{\'a}ros, P.\ 2016, \prd, 93, 083003. doi:10.1103/PhysRevD.93.083003
\bibitem[Shibata \& Taniguchi(2006)]{2006PhRvD..73f4027S} Shibata, M. \& Taniguchi, K.\ 2006, \prd, 73, 064027. doi:10.1103/PhysRevD.73.064027
\bibitem[Sun et al.(2017)]{2017ApJ...835....7S} Sun, H., Zhang, B., \& Gao, H.\ 2017, \apj, 835, 7. doi:10.3847/1538-4357/835/1/7
\bibitem[Thompson \& Duncan(1996)]{1996ApJ...473..322T} Thompson, C. \& Duncan, R.~C.\ 1996, \apj, 473, 322. doi:10.1086/178147
\bibitem[Thompson \& Duncan(1995)]{1995MNRAS.275..255T} Thompson, C. \& Duncan, R.~C.\ 1995, \mnras, 275, 255. doi:10.1093/mnras/275.2.255
\bibitem[Troja et al.(2007)]{2007ApJ...665..599T} Troja, E., Cusumano, G., O'Brien, P.~T., et al.\ 2007, \apj, 665, 599. doi:10.1086/519450
\bibitem[Typel et al.(2010)]{2010PhRvC..81a5803T} Typel, S., R{\"o}pke, G., Kl{\"a}hn, T., et al.\ 2010, \prc, 81, 015803. doi:10.1103/PhysRevC.81.015803
\bibitem[Usov(1992)]{1992Natur.357..472U} Usov, V.~V.\ 1992, \nat, 357, 472. doi:10.1038/357472a0
\bibitem[Virgili et al.(2011)]{2011ApJ...727..109V} Virgili, F.~J., Zhang, B., O'Brien, P., et al.\ 2011, \apj, 727, 109. doi:10.1088/0004-637X/727/2/109
\bibitem[Wanderman \& Piran(2015)]{2015MNRAS.448.3026W} Wanderman, D. \& Piran, T.\ 2015, \mnras, 448, 3026. doi:10.1093/mnras/stv123
\bibitem[Xie et al.(2020)]{2020ApJ...894...52X} Xie, W.-J., Zou, L., Liu, H.-B., et al.\ 2020, \apj, 894, 52. doi:10.3847/1538-4357/ab8302
\bibitem[Xue et al.(2019)]{2019Natur.568..198X} Xue, Y.~Q., Zheng, X.~C., Li, Y., et al.\ 2019, \nat, 568, 198. doi:10.1038/s41586-019-1079-5
\bibitem[Yu et al.(2010)]{2010ApJ...715..477Y} Yu, Y.-W., Cheng, K.~S., \& Cao, X.-F.\ 2010, \apj, 715, 477. doi:10.1088/0004-637X/715/1/477
\bibitem[Yuan et al.(2017)]{2017xru..conf..240Y} Yuan, W., Zhang, C., Ling, Z., et al.\ 2017, The X-ray Universe 2017, 240
\bibitem[Yuan et al.(2015)]{2015arXiv150607735Y} Yuan, W., Zhang, C., Feng, H., et al.\ 2015, arXiv:1506.07735. doi:10.48550/arXiv.1506.07735
\bibitem[Zhang(2013)]{2013ApJ...763L..22Z} Zhang, B.\ 2013, \apjl, 763, L22. doi:10.1088/2041-8205/763/1/L22
\bibitem[Zhang et al.(2024)]{2024GCN.35931....1Z} Zhang, W.~J., Mao, X., Zhang, W.~D., et al.\ 2024, GRB Coordinates Network, Circular Service, No. 35931, 35931
\bibitem[Zhang \& M{\'e}sz{\'a}ros(2001)]{2001ApJ...552L..35Z} Zhang, B. \& M{\'e}sz{\'a}ros, P.\ 2001, \apjl, 552, L35. doi:10.1086/320255
\bibitem[Zou et al.(2021)]{2021ApJ...921L...1Z} Zou, L., Zheng, T.-C., Yang, X., et al.\ 2021, \apjl, 921, L1. doi:10.3847/2041-8213/ac2ee4
\bibitem[Zou et al.(2019)]{2019ApJ...877..153Z} Zou, L., Zhou, Z.-M., Xie, L., et al.\ 2019, \apj, 877, 153. doi:10.3847/1538-4357/ab17dc
\bibitem[Zou \& Liang(2022)]{2022MNRAS.513L..89Z} Zou, L. \& Liang, E.-W.\ 2022, \mnras, 513, L89. doi:10.1093/mnrasl/slac040
\bibitem[Zou et al.(2021)]{2021MNRAS.508.2505Z} Zou, L., Liang, E.-W., Zhong, S.-Q., et al.\ 2021, \mnras, 508, 2505. doi:10.1093/mnras/stab2766


\end{thebibliography}

\clearpage
\begin{table}
\begin{center}
\caption{The fitting results for the different merger remnants with different EoSs}
\label{tab:fit}
\begin{tabular}{lcccccc}
\hline\hline

	&	$R_{0}$	($\rm Gpc^{-3}$ $\rm yr^{-1}$)		&	a			&	b			&	B			&	$\eta$			\\
\hline																					
DD2($M_{\rm TOV}=2.42M_{\odot}$)	&				&				&				&				&				\\
\hline																					
total	&	298.14 	$\pm$	4.66 	&	2.19 	$\pm$	0.03 	&	-4.45 	$\pm$	0.07 	&	5.74 	$\pm$	0.03 	&	-0.81 	$\pm$	0.03 	\\
stable NSs	&	214.15 	$\pm$	4.03 	&	2.13 	$\pm$	0.04 	&	-4.65 	$\pm$	0.09 	&	5.65 	$\pm$	0.03 	&	-0.77 	$\pm$	0.04 	\\
SMNSs	&	81.67 	$\pm$	1.18 	&	2.33 	$\pm$	0.03 	&	-4.07 	$\pm$	0.04 	&	6.01 	$\pm$	0.03 	&	-0.88 	$\pm$	0.03 	\\
HMNSs	&	2.50 	$\pm$	0.07 	&	2.15 	$\pm$	0.07 	&	-5.45 	$\pm$	0.23 	&	5.78 	$\pm$	0.07 	&	-0.57 	$\pm$	0.05 	\\
\hline																					
AB-L($M_{\rm TOV}=2.71M_{\odot}$)	&				&				&				&				&				\\
\hline																					
total	&	298.95 	$\pm$	4.68 	&	2.19 	$\pm$	0.03 	&	-4.45 	$\pm$	0.07 	&	5.74 	$\pm$	0.03 	&	-0.81 	$\pm$	0.03 	\\
stable NSs	&	291.63 	$\pm$	4.59 	&	2.19 	$\pm$	0.03 	&	-4.43 	$\pm$	0.06 	&	5.74 	$\pm$	0.03 	&	-0.82 	$\pm$	0.03 	\\
SMNSs	&	6.82 	$\pm$	0.19 	&	2.23 	$\pm$	0.07 	&	-5.41 	$\pm$	0.20 	&	5.89 	$\pm$	0.06 	&	-0.59 	$\pm$	0.05 	\\
HMNSs	&	0.57 	$\pm$	0.03 	&	2.29 	$\pm$	0.15 	&	-5.07 	$\pm$	0.36 	&	6.21 	$\pm$	0.13 	&	-0.57 	$\pm$	0.09 	\\
\hline																					
AB-N($M_{\rm TOV}=2.67M_{\odot}$)	&				&				&				&				&				\\
\hline																					
total	&	298.84 	$\pm$	4.68 	&	2.19 	$\pm$	0.03 	&	-4.45 	$\pm$	0.07 	&	5.74 	$\pm$	0.03 	&	-0.81 	$\pm$	0.03 	\\
stable NSs	&	290.84 	$\pm$	4.58 	&	2.19 	$\pm$	0.03 	&	-4.43 	$\pm$	0.06 	&	5.74 	$\pm$	0.03 	&	-0.82 	$\pm$	0.03 	\\
SMNSs	&	7.51 	$\pm$	0.22 	&	2.23 	$\pm$	0.07 	&	-5.40 	$\pm$	0.20 	&	5.89 	$\pm$	0.06 	&	-0.60 	$\pm$	0.05 	\\
HMNSs	&	0.57 	$\pm$	0.03 	&	2.25 	$\pm$	0.13 	&	-5.01 	$\pm$	0.33 	&	6.22 	$\pm$	0.12 	&	-0.61 	$\pm$	0.09 	\\
\hline																										GM1($M_{\rm TOV}=2.37M_{\odot}$)	&				&				&				&				&				\\
\hline																					
total	&	297.98 	$\pm$	4.67 	&	2.19 	$\pm$	0.03 	&	-4.45 	$\pm$	0.07 	&	5.74 	$\pm$	0.03 	&	-0.81 	$\pm$	0.03 	\\
stable NSs	&	17.54 	$\pm$	0.25 	&	2.25 	$\pm$	0.03 	&	-5.61 	$\pm$	0.10 	&	5.02 	$\pm$	0.02 	&	-0.68 	$\pm$	0.03 	\\
SMNSs	&	276.35 	$\pm$	4.62 	&	2.19 	$\pm$	0.04 	&	-4.42 	$\pm$	0.07 	&	5.79 	$\pm$	0.03 	&	-0.81 	$\pm$	0.03 	\\
HMNSs	&	3.71 	$\pm$	0.10 	&	2.09 	$\pm$	0.07 	&	-5.43 	$\pm$	0.21 	&	5.67 	$\pm$	0.06 	&	-0.59 	$\pm$	0.05 	\\
\hline																					
APR($M_{\rm TOV}=2.20M_{\odot}$)	&				&				&				&				&				\\
\hline																					
total	&	294.65 	$\pm$	4.62 	&	2.19 	$\pm$	0.03 	&	-4.44 	$\pm$	0.06 	&	5.74 	$\pm$	0.03 	&	-0.81 	$\pm$	0.03 	\\
SMNSs	&	290.30 	$\pm$	4.58 	&	2.19 	$\pm$	0.03 	&	-4.42 	$\pm$	0.06 	&	5.74 	$\pm$	0.03 	&	-0.82 	$\pm$	0.03 	\\
HMNSs	&	4.39 	$\pm$	0.17 	&	2.35 	$\pm$	0.09 	&	-5.45 	$\pm$	0.24 	&	6.03 	$\pm$	0.07 	&	-0.60 	$\pm$	0.06 	\\			
\hline																				
\hline																				\end{tabular}
\end{center}
\end{table}

\begin{table}
\begin{center}
\caption{X-ray transients discovered by observations with the {\it Chandra}/CDF-S, {\it Swift}/BAT, and {\it Swift}/XRT in our sample}
\label{tab:paras}
\begin{tabular}{lcccccc}
\hline\hline

XT  & $z$ &  $\Gamma$ & $\log L_{\rm b}$ (erg $\rm s^{-1}$) &  $\alpha_{1}$ &  $\alpha_{2}$ &  $t_{\rm b}/10^3\rm s$ \\
\hline
XT2	&	0.738	&	1.57 	$\pm$	0.55 	&	44.98 	$\pm$	0.05 	&	0.09 	$\pm$	0.11 	&	2.43 	$\pm$	0.19 	&	2.53 	$\pm$	0.24 	\\
XT1	&	2.23	&	1.43 	$\pm$	0.26 	&	47.21 	$\pm$	0.10 	&	-0.04 	$\pm$	0.10 	&	1.60 	$\pm$	0.10 	&	0.20 	$\pm$	0.10 	\\
080819	&	0.7	&	3.00 	$\pm$	1.10 	&	44.07 	$\pm$	0.17 	&	-0.20 	$\pm$	0.10 	&	2.80 	$\pm$	1.90 	&	5.30 	$\pm$	0.20 	\\
170901	&	1.44	&	2.20 	$\pm$	0.60 	&	45.86 	$\pm$	0.06 	&	0.10 	$\pm$	0.10 	&	1.90 	$\pm$	0.50 	&	2.10 	$\pm$	0.30 	\\
210423	&	1.51	&	2.20 	$\pm$	0.70 	&	45.68 	$\pm$	0.05 	&	0.20 	$\pm$	0.10 	&	3.80 	$\pm$	1.20 	&	4.40 	$\pm$	0.40 	\\

\hline\hline
sGRB-XP  & $z$ &  $\Gamma$ &  $\log L_{\rm b}$ (erg $\rm s^{-1}$) &  $\alpha_{1}$ &  $\alpha_{2}$ & $t_{\rm b}/10^3\rm s$ \\
\hline
051221A	&	0.5465	&	1.83 	$\pm$	0.21 	&	45.25 	$\pm$	0.10 	&	0.31 	$\pm$	0.11 	&	1.34 	$\pm$	0.07 	&	28.53 	$\pm$	6.92 	\\
060614	&	0.1254	&	1.75 	$\pm$	0.01 	&	44.43 	$\pm$	0.03 	&	0.05 	$\pm$	0.05 	&	1.82 	$\pm$	0.04 	&	45.73 	$\pm$	2.62 	\\
060801	&	1.31	&	1.64 	$\pm$	0.37 	&	48.34 	$\pm$	0.09 	&	-0.20 	$\pm$	1.16 	&	1.16 	$\pm$	0.27 	&	0.15 	$\pm$	0.07 	\\
061201	&	0.111	&	1.30 	$\pm$	0.18 	&	45.20 	$\pm$	0.17 	&	0.62 	$\pm$	0.12 	&	2.12 	$\pm$	0.13 	&	2.88 	$\pm$	0.64 	\\
070724A	&	0.457	&	1.70 	$\pm$	0.78 	&	44.24 	$\pm$	1.24 	&	0.20 	$\pm$	0.00 	&	2.53 	$\pm$	0.00 	&	41.22 	$\pm$	4.28 	\\
070809	&	0.219	&	1.48 	$\pm$	0.32 	&	44.55 	$\pm$	0.15 	&	0.04 	$\pm$	0.18 	&	1.17 	$\pm$	0.28 	&	7.11 	$\pm$	3.74 	\\
090426	&	2.6	&	1.81 	$\pm$	0.34 	&	48.40 	$\pm$	0.12 	&	-0.40 	$\pm$	0.88 	&	1.05 	$\pm$	0.11 	&	0.22 	$\pm$	0.09 	\\
100117A	&	0.92	&	1.45 	$\pm$	0.12 	&	48.35 	$\pm$	0.07 	&	0.67 	$\pm$	0.16 	&	4.53 	$\pm$	0.17 	&	0.23 	$\pm$	0.02 	\\
100625A	&	0.425	&	2.03 	$\pm$	0.32 	&	46.26 	$\pm$	0.24 	&	0.46 	$\pm$	0.59 	&	3.16 	$\pm$	1.02 	&	0.24 	$\pm$	0.07 	\\
101219A	&	0.718	&	1.34 	$\pm$	0.32 	&	47.94 	$\pm$	0.15 	&	0.23 	$\pm$	0.69 	&	6.24 	$\pm$	1.91 	&	0.17 	$\pm$	0.02 	\\
130603B	&	0.356	&	1.77 	$\pm$	0.20 	&	46.21 	$\pm$	0.11 	&	0.42 	$\pm$	0.07 	&	1.68 	$\pm$	0.07 	&	3.13 	$\pm$	0.60 	\\
140903A	&	0.351	&	1.60 	$\pm$	0.21 	&	45.72 	$\pm$	0.06 	&	0.07 	$\pm$	0.07 	&	1.22 	$\pm$	0.08 	&	8.52 	$\pm$	1.52 	\\
190627A	&	1.942	&	1.99 	$\pm$	0.16 	&	47.57 	$\pm$	0.09 	&	0.09 	$\pm$	0.11 	&	1.51 	$\pm$	0.09 	&	21.36 	$\pm$	3.88 	\\

\hline\hline

sGRB-EE  & $z$ & $\Gamma$ &   $\log L_{\rm b}$ (erg $\rm s^{-1}$) &  $\alpha_{1}$ &  $\alpha_{2}$ &  $t_{\rm b}/10^3\rm s$\\
\hline
050724	&	0.257	&	1.65 	$\pm$	0.17 	&	48.41 	$\pm$	0.15 	&	0.15 	$\pm$	0.13 	&	3.09 	$\pm$	0.07 	&	0.11 	$\pm$	0.00 	\\
061006	&	0.4377	&	1.74 	$\pm$	0.17 	&	48.66 	$\pm$	1.28 	&	0.34 	$\pm$	0.40 	&	4.90 	$\pm$	0.70 	&	0.06 	$\pm$	0.01 	\\
061210	&	0.41	&	1.56 	$\pm$	0.28 	&	48.04 	$\pm$	2.79 	&	0.18 	$\pm$	0.27 	&	1.60 	$\pm$	0.81 	&	1.20 	$\pm$	3.84 	\\
070714B	&	0.92	&	1.36 	$\pm$	0.19 	&	49.79 	$\pm$	0.03 	&	-0.16 	$\pm$	0.19 	&	2.41 	$\pm$	0.08 	&	0.09 	$\pm$	0.00 	\\
071227	&	0.383	&	1.91 	$\pm$	0.53 	&	47.42 	$\pm$	0.21 	&	0.58 	$\pm$	0.13 	&	6.69 	$\pm$	0.79 	&	0.18 	$\pm$	0.01 	\\
080123	&	0.495	&	1.50 	$\pm$	0.13 	&	48.83 	$\pm$	0.04 	&	0.25 	$\pm$	0.32 	&	1.92 	$\pm$	0.11 	&	0.07 	$\pm$	0.02 	\\
090510	&	0.903	&	1.60 	$\pm$	0.12 	&	47.58 	$\pm$	0.08 	&	0.70 	$\pm$	0.04 	&	2.26 	$\pm$	0.11 	&	1.65 	$\pm$	0.25 	\\
100724A	&	1.288	&	1.95 	$\pm$	0.27 	&	47.58 	$\pm$	0.10 	&	0.26 	$\pm$	0.44 	&	2.35 	$\pm$	0.40 	&	0.69 	$\pm$	0.16 	\\
111117A	&	2.211	&	1.50 	$\pm$	0.40 	&	48.31 	$\pm$	0.10 	&	0.10 	$\pm$	0.40 	&	3.10 	$\pm$	1.50 	&	0.18 	$\pm$	0.05 	\\
131004A	&	0.717	&	1.90 	$\pm$	0.50 	&	47.34 	$\pm$	0.32 	&	-0.20 	$\pm$	0.57 	&	1.61 	$\pm$	0.20 	&	0.84 	$\pm$	0.50 	\\
150423A	&	1.394	&	1.26 	$\pm$	0.23 	&	46.90 	$\pm$	0.09 	&	0.20 	$\pm$	0.44 	&	2.10 	$\pm$	0.50 	&	0.77 	$\pm$	0.13 	\\
\hline\hline
\end{tabular}
\end{center}
\end{table}

\begin{table}
\begin{center}
\caption{The event rates of XTs powered by the remnant types for different EoSs}
\label{tab:N}
\begin{tabular}{lcccccc}
\hline\hline
	&	stable NSs ($\rm yr^{-1}$)	&	SMNSs ($\rm yr^{-1}$)	&	HMNSs ($\rm yr^{-1}$)	\\
\hline									

DD2	&	1226.28 	&	539.75 	&	13.84 	\\
AB-L	&	1750.56 	&	40.84 	&	3.48 	\\
AB-N&	1745.82 	&	45.35 	&	3.52 	\\
GM1	&	100.57 	&	1660.44 	&	19.87 	\\
APR &       -               &      1740.51    &      29.21   \\
\hline									
\hline									
\end{tabular}
\end{center}
\end{table}

\clearpage

\end{document}